\documentclass[twocolumn,showpacs,showkeys,nofootinbib,superscriptaddress]{revtex4}

\usepackage{graphicx}
\usepackage{subfigure}
\usepackage{amsmath,amsfonts,amssymb,bm}

\begin{document}

\title{Scattering of $\mathbf{p}\bm{\mu}$ muonic atoms in solid hydrogen}

\author{J.~Wo\'zniak}
\affiliation{University of Mining and Metallurgy,
Fac.~Phys.~Nucl.~Techniques, PL--30059 Cracow, Poland}
\email{wozniak@ftj.agh.edu.pl}

\author{A.~Adamczak}
\affiliation{Institute of Nuclear Physics, PL--31342 Cracow, Poland}

\author{G.A.~Beer}
\affiliation{University of Victoria, Victoria, V8W 2Y2, Canada}

\author{V.M.~Bystritsky}
\affiliation{Joint Institute for Nuclear Research, Dubna 141980,
Russia}

\author{M.~Filipowicz}
\affiliation{University of Mining and Metallurgy, Fac.~of Fuels and
Energy, PL--30059 Cracow, Poland}

\author{M.C.~Fujiwara}
\affiliation{Department of Physics, University of Tokyo, Tokyo,
113--0033, Japan}

\author{T.M.~Huber}
\affiliation{Gustavus Adolphus College, St.Peter, MN 56082, USA}

\author{O.~Huot}
\affiliation{Department of Physics, University of Fribourg, CH--1700
Fribourg, Switzerland}

\author{R.~Jacot-Guillarmod}
\affiliation{Department of Physics, University of Fribourg, CH--1700
Fribourg, Switzerland}

\author{P.~Kammel}
\affiliation{University of Illinois at Urbana--Champaign, Urbana, IL
61801, USA}

\author{S.K.~Kim}
\affiliation{Physics Department, Joenbuk National University, Jeonju,
Jeonbuk 561--756, S.~Korea}

\author{P.E.~Knowles}
\affiliation{Department of Physics, University of Fribourg, CH--1700
Fribourg, Switzerland}

\author{A.R.~Kunselman}
\affiliation{University of Wyoming Physics, Laramie, WY 82071--3905,
USA}

\author{G.M.~Marshall}
\affiliation{TRIUMF, Vancouver, V6T 2A3, Canada}

\author{F.~Mulhauser}
\affiliation{Department of Physics, University of Fribourg, CH--1700
Fribourg, Switzerland}

\author{A.~Olin}
\affiliation{TRIUMF, Vancouver, V6T 2A3, Canada}

\author{C.~Petitjean}
\affiliation{Paul Scherrer Institute, CH--5232 Villigen, Switzerland}

\author{T.A.~Porcelli}
\affiliation{Department of Physics, University of Northern British
Columbia, Prince George, V2N 4Z9, Canada}

\author{L.A.~Schaller}
\affiliation{Department of Physics, University of Fribourg, CH--1700
Fribourg, Switzerland}

\author{V.A.~Stolupin}
\affiliation{Joint Institute for Nuclear Research, Dubna 141980,
Russia}

\author{J.~Zmeskal} 
\affiliation{Institute for Medium Energy Physics, Austrian Academy of
Sciences, A--1090 Wien, Austria}

\collaboration{TRIUMF Muonic Hydrogen Collaboration}
\noaffiliation


\begin{abstract}  
We present the results of experimental and theoretical study of the
scattering of low energy $\mathrm{p}\mu$ atoms in solid hydrogen cooled to
3~K\@.
Strong effects resulting from the solid state interactions have been
observed in the TRIUMF experiment E742 where muons were stopped in
thin frozen layers of hydrogen.
The resulting emission of low energy $\mathrm{p}\mu$ atoms from the hydrogen
layer into the adjacent vacuum was much higher than that predicted by
calculations which ignored the solid nature of the hydrogen.
New differential scattering cross sections have been calculated for
the collisions of $\mathrm{p}\mu$ atoms on solid hydrogen to account for its
quantum crystalline nature.
Analysis of the experimental data performed using such cross sections
shows the important role of the coherent scattering in $\mathrm{p}\mu$ atom
diffusion.
For $\mathrm{p}\mu$ energies lower than the Bragg cutoff limit
($\approx 2$~meV) the elastic Bragg scattering vanishes which makes
the total scattering cross section fall by several orders of
magnitude, and thus the hydrogen target becomes transparent allowing
the emission of cold $\mathrm{p}\mu$ atoms to occur.
\end{abstract}


\maketitle

\section{Introduction}
\label{sec:introduction}

Negative muons stopping in hydrogen can form muonic hydrogen
($\mathrm{p}\mu$) atoms.
Although created in excited states, such atoms cascade to the ground
state quickly ($10^{-12}$~s) where their kinetic energy is of the
order of several eV, much higher than thermal equilibrium energies.
The muonic hydrogen atom is about 200 times smaller (m$_\mu$/m$_e$
scaling) than the size of ordinary electronic hydrogen.
The small neutral atom can easily diffuse through the surrounding
medium undergoing different types of interaction including elastic and
inelastic scattering.
Scattering of fast $\mathrm{p}\mu$ atoms in hydrogen is governed by a large
cross section ($\sigma_s > 10^{-19}\:\mathrm{cm}^2$) which is quite
effective at slowing them down.

Only a few experiments have examined the scattering of muonic atoms on
nuclei and molecules directly, although it is an important process in
most muon physics phenomena such as muon catalyzed nuclear
fusion ($\mu$CF) or muon nuclear capture by protons (see
reviews~\cite{breun89,ponom90,froel92,measd01}).
The first experiments of $\mathrm{p}\mu +\mathrm{H}_2$ scattering were
performed in gaseous hydrogen and used the traditional diffusion
method~\cite{dzhel66,berti75,berti82,bystr84}.

Much more has been done to study the scattering theoretically.
Many calculations of cross sections for scattering on bare nuclei,
atoms, and hydrogen molecules have been made, however, solid
state effects were not considered.
The high accuracy calculations of the total cross sections for low
energy scattering ($\varepsilon_{coll}<50$~eV) for $\mathrm{p}\mu$ and
other muonic atoms on bare hydrogen nuclei (called the ``nuclear''
cross sections) were done in Ref.~\cite{bracc89b} by solving the
Coulomb three--body scattering problem using the adiabatic
multichannel approach.
Differential cross sections for that case were calculated in
Ref.~\cite{melez92} using phase shift values from
Ref.~\cite{bracc89b}.
For collision energies lower than about $0.1-1$~eV it is necessary to
account for both electron screening and the molecular structure of the
target.
Total and differential cross sections for this case (called the
``gas'' cross sections) are given in Ref.~\cite{adamc96} and in
Ref.~\cite{adamc93}, respectively.
Another possible approach to including the molecular effects for
epithermal energies uses the Sachs--Teller tensor--of--mass model and
can be found in Refs.~\cite{cohen86,bouko96c}.

The scattering experiment results given in
Refs.~\cite{dzhel66,berti75,berti82,bystr84} were sometimes
inconsistent not only among themselves but with theory as well
(see~\cite{bystr95} for a review).
The latest and most advanced measurements in gaseous hydrogen were
performed at PSI~\cite{abbot97} where the cross sections for
$\mathrm{p}\mu$ scattering on $\mathrm{H}_2$ molecules~\cite{adamc96}
were used for the analysis of the experimental data.
Those measurements were not in full agreement with the theory.  
On the other hand, the $\mathrm{d}\mu + \mathrm{D}_2$ scattering
measurements performed by the same collaboration~\cite{abbot97} were
in agreement.

Until now, no experimental studies concerning $\mu$--atom scattering
in solid hydrogen have been performed.
Such experiments are complicated to analyze because the results for
the cross section are not directly obtained but are only deduced by
their effect on other results, such as time distributions or yield
intensities, which themselves are often obscured by other background
processes.

The development at TRIUMF of the multilayer thin frozen hydrogen film
targets~\cite{marsh93b,knowl96,mulha96,knowl97,fujiw00,porce01,marsh01},
which produce muonic atom beams emitted into vacuum, permitted the
cross sections to be probed in another way.
We have studied several isolated muon induced processes using a
time--of--flight (TOF) method permitted by the frozen target
geometry~\cite{marsh01}.
In particular it was used in TRIUMF experiment E742 for the cross
section study of $\mathrm{d}\mu+\mathrm{H}_2$,
$\mathrm{t}\mu+\mathrm{H}_2$ scattering and the Ramsauer--Townsend
effect (RT) which is seen for these systems at collision energies
between $2-10$~eV~\cite{jacot96,mulha99,mulha01}.
During those measurements a strong emission of low energy
$\mathrm{p}\mu$ atoms from the hydrogen layers into adjacent vacuum
was observed.
The yield was much higher than expected based on calculations which
ignored the solid nature of the hydrogen target.
Additional experimental studies~\cite{wozni99,bystr01} and new
theoretical calculations of ``solid'' cross sections~\cite{adamc99}
have been performed in order to clarify and explain the
$\mathrm{p}\mu$ emission behavior.

This paper summarizes our findings.
In section II the theoretical background and new calculations of
scattering cross sections in solid hydrogen are described.
The experimental apparatus and the measurement method are given in
section III\@.
Section IV presents the measurement results and their analysis,
whereas section V contains the discussion of the results and some
concluding remarks.

\section{Scattering cross sections}
\label{sec:scatt-cross-sect}

\subsection{Scattering on bare nuclei, atoms and molecules}
\label{sec:scatt-bare-nucl}

\begin{figure}[t]
\includegraphics[width=0.48\textwidth]{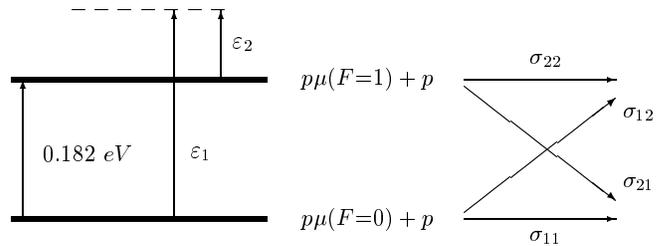} 
\caption{Energy levels for the $\mathrm{p}\mu+\mathrm{p}$ system,
where $\varepsilon_1$ and $\varepsilon_2$ represent the collision
energies for the singlet ($\mathrm{F}=0$) and triplet ($\mathrm{F}=1$)
states, respectively.
The dashed line represents the total energy of the system.
Four transitions with cross sections $\sigma_{ik}$, $i,k=1,2$ are
possible due to the hyperfine splitting of the energy levels.
For the scattering of $\mathrm{p}\mu$ in the singlet state only the
elastic scattering $\sigma_{11}$ is possible if the collision energy
$\varepsilon _1$ is below $\mathrm{E}_{hfs} = 0.182\:\mathrm{eV}$\@.}
\label{fig:diagram}
\end{figure}

Cross section calculations for the 1S state $\mathrm{p}\mu$ atom
scattering on bare protons
\begin{equation}
{\mathrm p}\mu({\mathrm F}) + {\mathrm p} \rightarrow 
{\mathrm p}\mu({\mathrm F}') + {\mathrm p},
\label{eq1}
\end{equation}
(where $\mathrm{F}$, $\mathrm{F}'$ are the initial and final muonic
atom spins) were begun by Gershtein~\cite{gersh58}.
He treated the process as a quantum mechanical Coulomb three--body problem.
Figure~\ref{fig:diagram} shows the diagram of the two isolated states
of the $\mathrm{p}\mu+\mathrm{p}$ system with the possible
transitions.

Most of the following calculations were performed in the adiabatic
representation which results from expanding the wave function of a
three--body system over a complete set of solutions of the two--center
Coulomb problem~\cite{vinit82}, especially in the two--level
approximation~\cite{matve75} with further
modifications~\cite{ponom79,bubak87,strue86,cohen91}.
The progress in the effective potential calculation for the two--center
problem permits the multichannel scattering equations to be solved
even when there are a large number of closed
channels~\cite{melez83,melez86}.
Accurate calculations of the total cross sections performed in this 
multichannel approach including $\mathrm{p}\mu+\mathrm{p}$ scattering 
are presented in Ref.~\cite{bracc89b}.
Reactance T--matrices and phase shifts, also given in
Ref.~\cite{bracc89b}, for different values of the total orbital
angular momentum and the total spin of the three--particle system have
been used to calculate the differential cross sections~\cite{melez92}.
Cross sections (total and differential) for collisions with energies
less than $\approx 1\:\mathrm{eV}$, where both electron screening and
molecular binding are important (``gas'' cross sections), are given in
Refs.~\cite{adamc96,adamc93}.  
The screening effect is described there in terms of the effective
screening potential, and the Fermi pseudopotential method was applied
to model the chemical binding.

\subsection{Cross sections for solid hydrogen}
\label{sec:cross-sections-solid}

The TRIUMF
experiments~\cite{marsh93b,knowl96,mulha96,knowl97,fujiw00,porce01,marsh01}
have stimulated theoretical studies of $\mu$--atom scattering in
solids.
Solid hydrogen at zero pressure is a quantum molecular crystal, which is
characterized by a~large amplitude of zero--point vibrations of the
molecules.
At 3~K, the vibration amplitude is approximately 18\% of the nearest
neighbor distance for the $\mathrm{H}_2$ molecule, and 15\% for the
$\mathrm{D}_2$ molecule~\cite{silve80}.
Nevertheless, experiments show that quantum solids display typical
crystal structures and that common crystal characteristics, such as
the density of the vibrational states, are well defined.
This proves that the molecular motion is correlated in such a manner
that the crystalline structure is not destroyed.
However, theoretical methods developed for a classical--crystal
description, encounter certain problems when applied in the case of
quantum crystals.
Namely, the interaction potential between the hydrogen molecules has a
highly repulsive anharmonic core and thus the standard lattice
dynamics leads to imaginary vibration frequencies.
Nevertheless, the standard dynamics can be used, after a
renormalization of the interaction potential by accounting for the
short--range correlations between neighboring--molecule movement.

Solid hydrogen can exist in both efficient--packing structures:
face--centered cubic (\textit{fcc}) and/or hexagonal close packed
(\textit{hcp}).
Since the TRIUMF targets were formed by rapidly freezing hydrogen gas on
a gold foil at 3~K and zero pressure, the solid layer has a
polycrystalline \textit{fcc} structure~\cite{silve80}.
There is also experimental evidence that thin hydrogen films formed on
\textit{fcc} metal (e.g., gold or silver) will retain that same
structure~\cite{souer86}.
Since the \textit{fcc} and \textit{hcp} crystals are very similar
(e.g., the molar volumes are almost the same and the first three
shells of neighbors of any fixed molecule are identical in both the
structures), the \textit{fcc} cross sections are also a good
approximation to the \textit{hcp} case.
The hydrogen was purified by a palladium filter at 600~K immediately
prior to freezing, so the resulting solid target had a statistical
distribution (1:3) of molecular rotational states $K=0$ and $K=1$.
Such a mixture of rotational states is often called ``normal''
hydrogen, $\mathrm{nH}_2$.
The relevant lattice constant for the \textit{fcc} structure at zero
pressure is 0.5338~nm~\cite{silve80}.

A method to calculate the scattering cross sections of muonic hydrogen
atoms in solid hydrogen (``solid'' cross sections) based on Van Hove's
approach and using phase shifts for muonic atom scattering on bare
nuclei~\cite{bracc89b,chicc92} has been proposed by Adamczak
(see~\cite{adamc99} for details).
The calculated differential cross sections include incoherent and
coherent effects.
The impinging muonic atom can induce inelastic reactions, both in a
single molecule (rovibrational transitions, or spin--flip) and in the
whole target (excitations or deexcitations of the lattice vibrational
states).
The latter are usually interpreted as creation or annihilation of
phonons.
It is possible to create or annihilate one or more phonons in coherent
or incoherent processes, but in practice, annihilation processes are
strongly suppressed in a 3~K target because few phonons exist at low
temperatures.

\begin{figure}[t] 
  \includegraphics[width=0.48\textwidth]{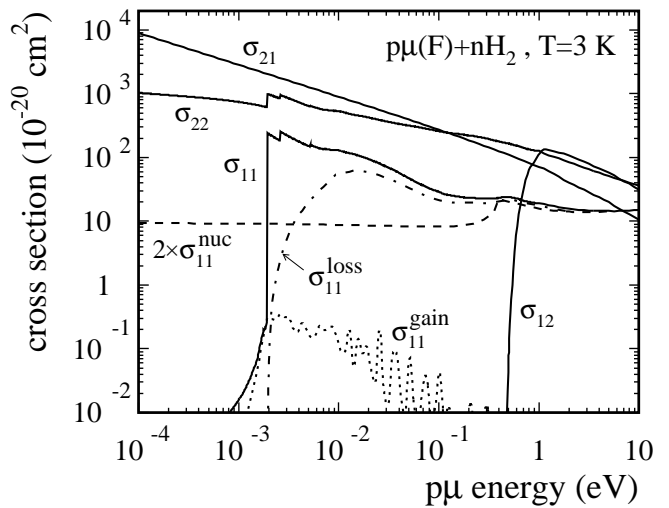}
  \caption{Total cross sections for $\mathrm{p}\mu (\mathrm{F})$
    scattering in 3--K polycrystalline nH$_2$ with the \textit{fcc}
    structure, for different values of the initial and final muonic
    atom spin, $\mathrm{F}$. The dotted line represents the
    phonon--annihilation fraction of $\sigma_{11}$ that results in
    $\mathrm{p}\mu$ energy gain; the sum of contributions from phonon
    creation and rovibrational excitations to $\sigma_{11}$, which
    lead to $\mathrm{p}\mu$ energy loss, is denoted by dash--dotted
    line. The doubled nuclear scattering cross section
    $\sigma_{11}^{\mathrm{nuc}}$ for
    $\mathrm{p}\mu(\mathrm{F}=0)+\mathrm{p}$ is shown for comparison
    (dashed line). Note the Bragg cutoff energy, $\mathrm{E}_B$, at 
    $\varepsilon \approx 2$~meV for $\sigma_{11}$.
    \label{fig:xppp-all}}
\end{figure}

Figure~\ref{fig:xppp-all} presents the calculated total cross
sections for $\mathrm{p}\mu$ scattering on 3--K solid \textit{fcc}
hydrogen for different initial and final spin states $\mathrm{F}$ of
the $\mathrm{p}\mu$ atom.
Also are shown some details of the total cross section for
$\mathrm{p}\mu$ atom scattering from the ground spin state
$\mathrm{F}=0$.
For the sake of comparison, the doubled cross section
$\sigma_{11}^{\mathrm{nuc}}$ of $\mathrm{p}\mu(\mathrm{F}=0)+\mathrm{p}$
nuclear scattering is plotted.
The cross sections are given for a single bound molecule. 
At energies greater than roughly 1~eV both the solid state and
molecular binding effects are very small and therefore the cross
section for a real hydrogen target is well described by the nuclear
cross section.

In Fig.~\ref{fig:xppp-all}, there is an important difference between
the singlet ($\sigma_{11}$, $\mathrm{F}=0$) and triplet
($\sigma_{22}$, $\mathrm{F}=1$) state scattering.
For the singlet, only the state $\mathrm{J}=\tfrac{1}{2}$ of the total
spin of the $\mathrm{p}\mu+\mathrm{p}$ system is possible.
As a result, the scattering in nH$_2$ is almost fully coherent and thus
interference effects determine the behavior of the singlet cross section
at the lowest energies.
Below the Bragg cutoff energy, $\mathrm{E}_B\approx{}2\:\mathrm{meV}$,
elastic and phonon--creation coherent scattering is impossible and the
total cross section is determined by the weak incoherent processes,
which gives rapid falloff of $\sigma_{11}$.
Coherent phonon annihilation is present below $\mathrm{E}_B$, but its
magnitude is very small at 3~K\@.
The rotational deexcitation $K=1\to K'=0$ of an $\mathrm{H}_2$
molecule gives no contribution to the cross section since this
transition is strictly forbidden for $\mathrm{F}=0$.

Scattering of the $\mathrm{p}\mu(\mathrm{F}=1)$ atom is possible in
the two total--spin states: $J=\tfrac{1}{2}$ and $J=\tfrac{3}{2}$.
The nuclear amplitudes for these two processes are very
different~\cite{bracc89b} and therefore averaging the molecular
scattering amplitude over spins leads to a strong incoherent
component.
As a result, the cross section $\sigma_{22}$ in solid hydrogen has a
large magnitude at $\varepsilon\to{}0$, though a small falloff of its
value at $\mathrm{E}_B$ is still present.
Significant contribution to $\sigma_{22}$ at lowest energies comes
also from the rotational deexcitation $K=1\to K'=0$, which is possible
(for $\mathrm{F}=1$) due to the exchange of the muon between the
protons during the collision process.

In Fig.~\ref{fig:xppp-all} are shown contributions to $\sigma_{11}$
from different processes.
The energy region $2-10$~meV is dominated by the strong elastic Bragg
scattering.
Phonon annihilation, denoted by the label
$\sigma_{11}^{\mathrm{gain}}$, is the only mechanism of
$\mathrm{p}\mu$ acceleration.
Weak incoherent elastic scattering is most important below
$\mathrm{E}_B$.
Slowing of $\mathrm{p}\mu$ is possible through the lattice excitations
and then, at sufficient incident energy, through subsequent rotational
and vibrational excitations.
The rovibrational transitions may take place with simultaneous one or
multiphonon creation.
The curve which shows the sum of contributions from all these
processes is labeled by $\sigma_{11}^{\mathrm{loss}}$.
At $\varepsilon\gtrsim\omega_D$, the inelastic processes are most
important and the cross section for the solid (per single molecule)
approaches the corresponding one for a free $\mathrm{H}_2$ molecule.

\begin{figure}[ht] 
\includegraphics[width=0.48\textwidth]{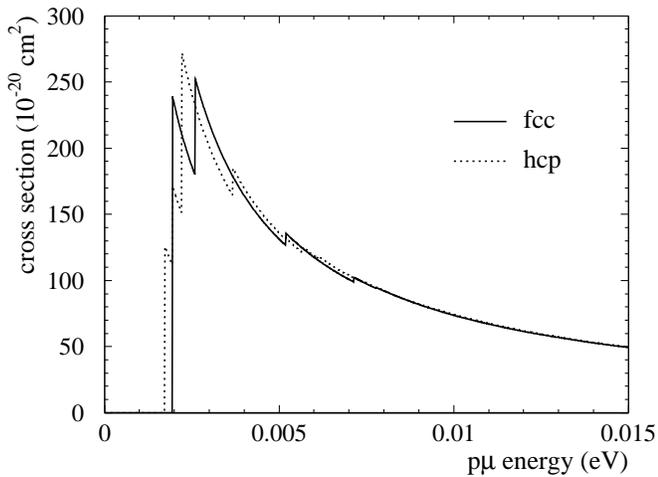}
\caption{Total cross section for Bragg scattering of
  $\mathrm{p}\mu(\mathrm{F}=0)$ atoms in 3--K polycrystalline nH$_2$
  with the \textit{fcc} (solid line) and \textit{hcp} (dotted line)
  structure.}
\label{fig:xbragg_hf}
\end{figure}
Figure~\ref{fig:xbragg_hf} illustrates the small differences between
Bragg scattering of $\mathrm{p}\mu$ in the \textit{fcc} and
\textit{hcp} polycrystalline nH$_2$.
The Bragg cutoff energy is slightly ($\approx 0.2$~meV) smaller in the
\textit{hcp} target.
Different Bragg peak patterns in the total cross sections are distinct
only below a few meV\@.
The magnitudes of the cross sections are similar in the two lattices
and, therefore, the theoretical estimation of cold--$\mathrm{p}\mu$
emission, obtained in this work for an~\textit{fcc} target, is also
valid in the \textit{hcp} case.

\subsection{Slowing down of $\mathrm{p}\mu$ atoms in solid hydrogen}
\label{sec:slowing-down-pmu}

The slowing down of $\mathrm{p}\mu$ in solid hydrogen has been
simulated by the Monte Carlo method using the new cross sections.
The more important characteristics are shown in the following figures.

\begin{figure}[t]
\includegraphics[width=0.48\textwidth]{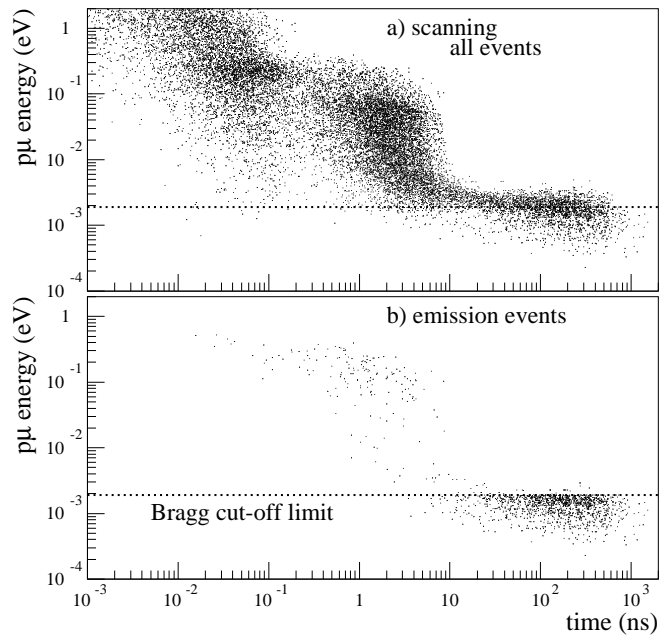}
\caption{Slowing down of $\mathrm{p}\mu$ atoms in a solid hydrogen
layer of thickness $3.4\:\mbox{mg}\cdot\mbox{cm}^{-2}$
(1000~Torr$\cdot \ell$, as defined in Section~\ref{sec:apparatus}).
The top plot a) shows the time and energy after each scattering event
during the whole slowing down process.
There are 2376~emission events and each $\mathrm{p}\mu$ atom undergoes
on the average 40~collisions before emission take place.
The bottom plot b) shows the time and energy for the $\mathrm{p}\mu$
atoms which have been emitted from the hydrogen layer.
The ``solid'' cross sections were used in the simulations.}
\label{fig:slow1}
\end{figure}

The simulations, performed with the Monte Carlo~\cite{wozni96},
represent a real experimental situation where muons were stopped and
formed $\mathrm{p}\mu$ atoms in a solid protium target of thickness
$3.4\ \mbox{mg}\cdot\mbox{cm}^{-2}$ (an experiment labeled later as
\#3 in Table~\ref{tab:useful}).
For this presentation only the histories of $\mathrm{p}\mu$ resulting in
the upstream emission from the hydrogen layer have been chosen.

\begin{figure}[t]
\includegraphics[width=0.48\textwidth]{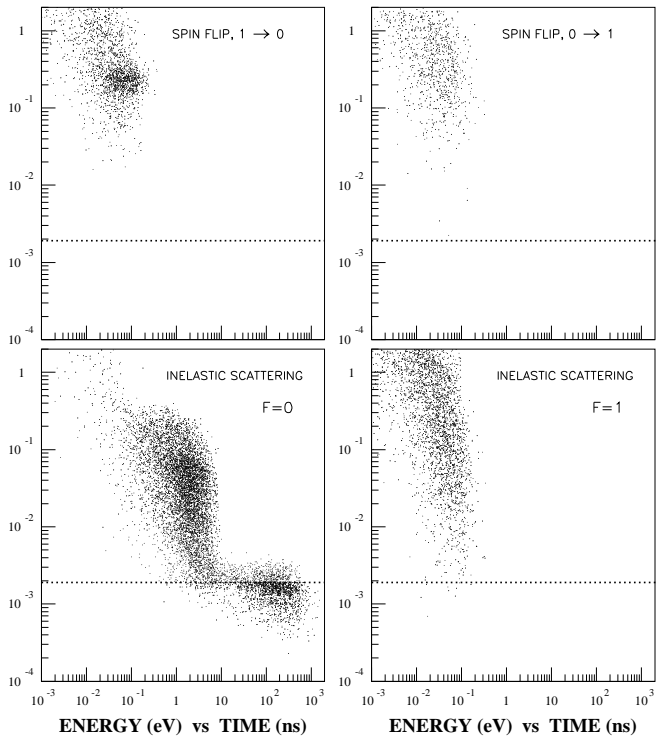}
\caption{Details of the scatterplot from Fig.~\ref{fig:slow1}a),
namely the spin--flip contribution (top) and the non--Bragg scattering
processes in the slowing down of $\mathrm{p}\mu$ atoms (bottom).
Note that downward spin--flip ($\mathrm{F}=1 \rightarrow
\mathrm{F}=0$) increases the energy by $\varepsilon_{hfs}= 0.182$~eV
and the upward spin--flip decreases the energy by that amount.
Elastic coherent scattering (not shown in the figure) is a dominant
process for energies near the Bragg cutoff energy limit ($\sim
2\cdot10^{-2}$~eV) and leaves the energy unchanged.}
\label{fig:slow2}
\end{figure}

Scatterplots on Figs.~\ref{fig:slow1} and \ref{fig:slow2} illustrate
the time dependence of the $\mathrm{p}\mu$ slowing down (shown are
2376~histories ending by the emission from the hydrogen layer).
Each point represents the $\mathrm{p}\mu$ energy after a scattering
event at a given time.
Figure~\ref{fig:slow1}a) shows a sampling of all events during slowing
down, whereas Fig.~\ref{fig:slow1}b) shows only the final coordinates
when $\mathrm{p}\mu$ emission has occurred.
One sees that the slowing down process is very fast and that after
approximately 10~ns the $\sim$meV energy region is reached.
Further decelerations of the $\mathrm{p}\mu$ are then slower
since the responsible inelastic cross sections become lower.
This transient region extends to $100-200$~ns when equilibrium energy
is reached and $\mathrm{p}\mu$ atom diffusion in hydrogen takes place.
The equilibrium energy is established near the Bragg cutoff limit
where
%
both the phonon creation and annihilation components of $\sigma_{11}$
become equal (see Fig.~\ref{fig:xppp-all}).

\begin{figure}[t]
\includegraphics[width=0.48\textwidth]{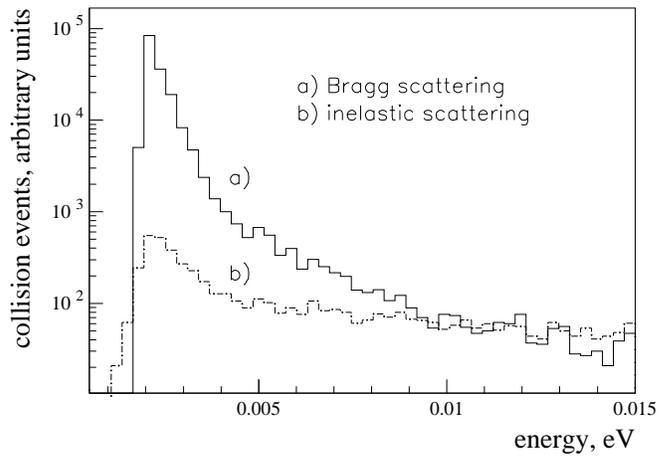}
\caption{Competition between Bragg scattering a) and non--Bragg
scattering b) in $\mathrm{p}\mu$ collisions in solid hydrogen for low
energies.  Note the logarithmic scales.}
\label{fig:sslow4}
\end{figure}

Figure~\ref{fig:slow2} represents contributions to slowing down from
separate processes for $\mathrm{p}\mu$ in singlet and triplet states.
In any experiment both $\mathrm{p}\mu$ atomic spin states will be
initially populated, however, the downward spin--flip is so fast in the
solid target that, after 0.1~ns, practically all $\mathrm{p}\mu$ atoms
are in the ground spin state.
Therefore, further slowing down is governed by the cross section
$\sigma_{11}$ (Fig.~\ref{fig:xppp-all}).
Efficient slowing down finishes after about 10~ns and subsequent
$\mathrm{p}\mu$ diffusion is determined by elastic Bragg scattering
and inelastic phonon scattering.
The competition between those two processes is shown in
Fig.~\ref{fig:sslow4}.

The elastic Bragg scattering does not change the $\mathrm{p}\mu$ kinetic
energy in a solid hydrogen (contrary to a gas, where the elastic scattering 
is an effective deceleration process because the $\mathrm{p}\mu$ atom can 
always transfer a part of its energy to a recoiling free H$_2$ molecule).
Only inelastic scattering can cause $\mathrm{p}\mu$ deceleration (or
acceleration from the phonon annihilation process) but is a weak
contribution at low energies and at low temperatures. 
Therefore, $\mathrm{p}\mu$ atoms spend a relatively long time in the 
diffusion stage before reaching the Bragg cutoff energy. 
In the case of a solid H$_2$, rapid 
falloff of the phonon creation cross section at energy $E_B$ makes the 
$\mathrm{p}\mu$ thermalisation less deep than in a gaseous hydrogen. 
Indeed, the equilibrium energy defined by 
the intersection of $\sigma_{11}^{\mathrm{gain}}$ and
$\sigma_{11}^{\mathrm{loss}}$ at $\approx 2$~meV 
(see Fig.~\ref{fig:xppp-all})
is still higher than thermal equilibrium in a 3--K gaseous H$_2$\@.
\begin{figure}[t]
\includegraphics[width=0.48\textwidth]{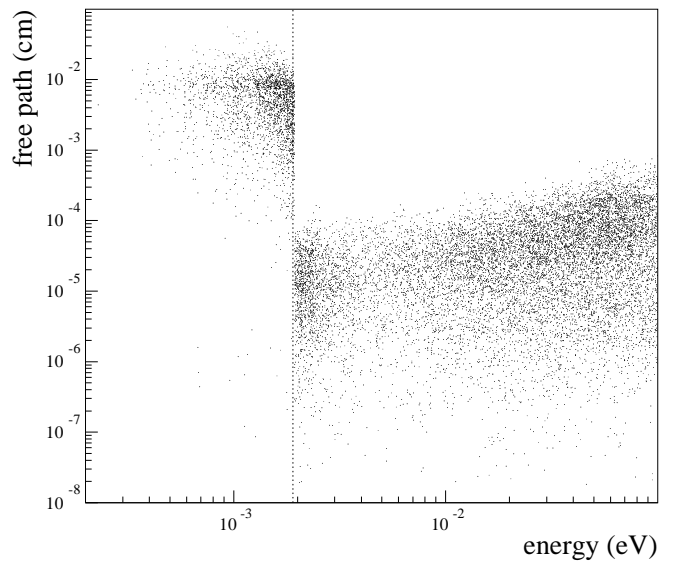}
\caption{Sampled values of the $\mathrm{p}\mu$ atom mean free path
between consecutive collisions versus the collision energy.
A strong increase of the mean free path is seen below the Bragg cutoff
energy. }
\label{fig:slow3}
\end{figure}

The strong increase of the $\mathrm{p}\mu(\mathrm{F}=0)$ atom mean
free path due to the sharp decrease of the cross section below the
Bragg cutoff energy is shown in Fig.~\ref{fig:slow3}.
Such behavior leads to an enhanced emission of cold $\mathrm{p}\mu$'s
from the thin solid hydrogen layers.
We note that a similar phenomenon is used in neutron physics to
extract cold neutrons from beams produced in nuclear reactors
(polycrystalline filters, see e.g.,~\cite{egels65}).

\section{Description of the experiment}
\label{sec:descr-exper}

\subsection{The apparatus}
\label{sec:apparatus}

\begin{figure}[t]
\begin{center}
\centerline{\setlength{\unitlength}{1mm}}
\begin{picture}(240,250)(0,0)
\put(40,-10){\includegraphics[height=9.5cm]{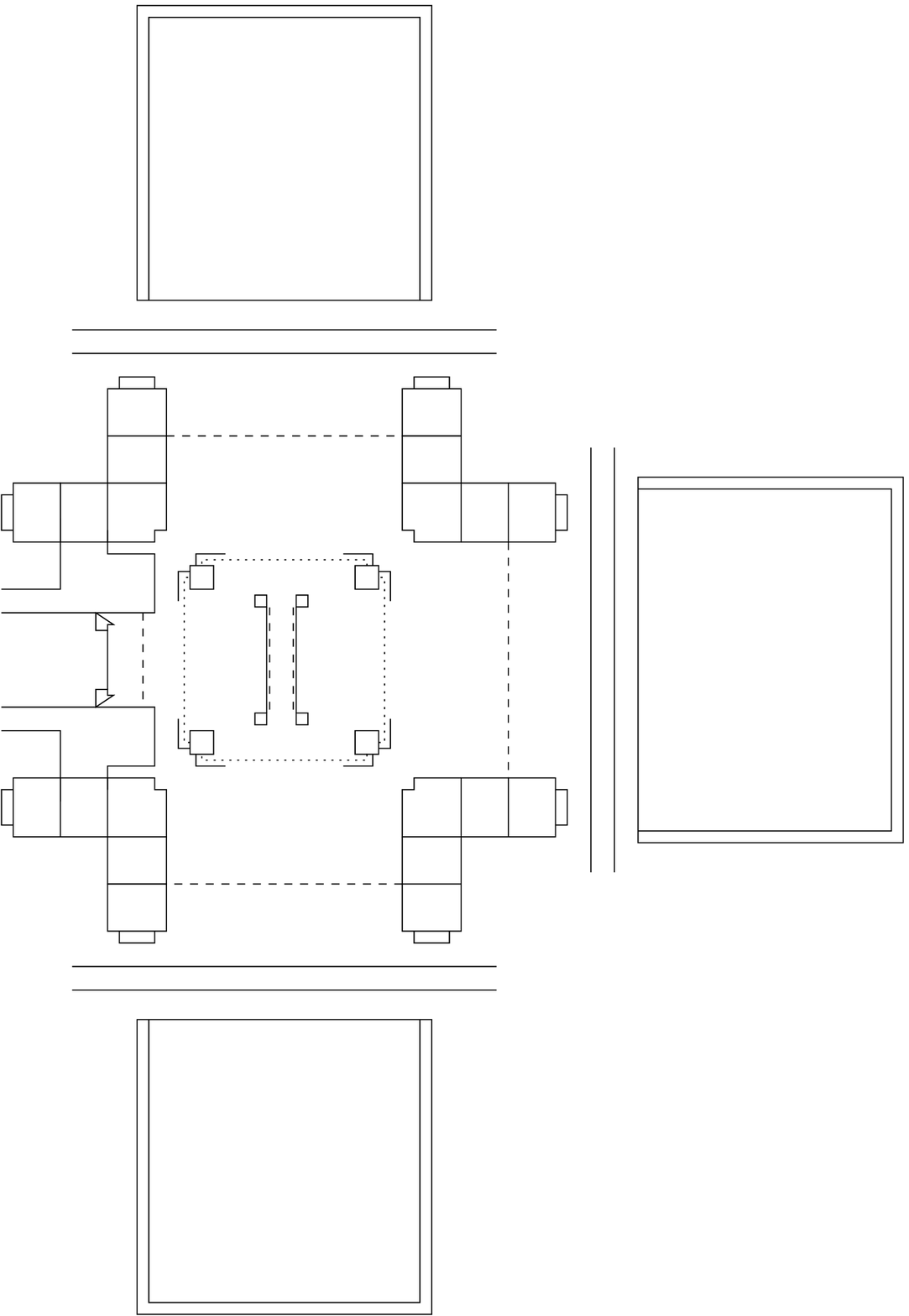} }
\put(52,127){\mbox{1}}
\put(20,125){\mbox{$\mu^-$}}
\put(30,125){\vector(1,0){30}}
\put(89,125){\mbox{2}}
\put(104,125){\mbox{3}}
\put(95,235){\mbox{(4)}}
\put(95,220){\mbox{G1}}
\put(85,210){\mbox{(75~cm$^3$)}}
\put(94,35){\mbox{(5)}}
\put(93,20){\mbox{G2}}
\put(81,10){\mbox{(122~cm$^3$)}}
\put(192,140){\mbox{(6)}}
\put(190,125){\mbox{NaI}}
\put(180,115){\mbox{(15~cm $\phi$}}
\put(180,105){\mbox{$\times$ 10~cm)}}
\put(148,188){\mbox{7}}
\put(163,175){\mbox{7}}
\put(148,58){\mbox{7}}
\end{picture}
\end{center}
\caption{The apparatus layout showing the muon entrance scintillator
  (1), the upstream (2) and downstream (3) gold foils (themselves
  inside the cryostat).
The surrounding detectors were the germanium detectors G1 (4) and G2
(5), the NaI (6), and the three pairs of electron counters (7).
The drawing is not strictly to scale.}
\label{fig:layout}
\end{figure}

The experiments studying $\mathrm{p}\mu$ scattering in solid hydrogen
were performed at the M20B muon channel at TRIUMF\@.
The layout of the apparatus is shown schematically in
Fig.~\ref{fig:layout}.
Gaseous hydrogen (or neon) was sprayed, using a special diffusion
system, onto the $51\:\mu\mbox{m}$ thick gold foil, maintained at 3~K,
where it froze creating the thin solid films which could be maintained
in high vacuum.
The diffuser was inserted from below and could be used to deposit gas
on either of the the two gold foils separately.
The thickness of the film was controlled by adjusting the
amount of gas injected.
Multi--layered targets could be made in which the frozen material
consisted of uniform layers, each made from different hydrogen
isotopes or other gases such as neon.
A versatile gas handling system allowed mixtures of different hydrogen
isotopes to be prepared with high precision.
The frozen film deposition uniformity has been measured independently
via energy loss of alpha particles~\cite{fujiw97b}.
The amount of gas injected into the system was conveniently measured
in units of $\mbox{Torr}\cdot\ell$, where the conversion factor
between $\mbox{Torr}\cdot\ell$ and $\mu\mbox{g}\cdot\mbox{cm}^{-2}$
has been determined~\cite{fujiw97b} and is on average
$3.4\:\mu\mbox{g}\cdot\mbox{cm}^{-2}$ per $\mbox{Torr}\cdot\ell$ for
$\mathrm{H}_2$.
Details of the target construction and working procedure are given in
Refs.~\cite{knowl96,marsh96}.
Details of the data acquisition electronics can be found in
Ref.~\cite{knowl97}.

Incident muons of momentum 26.70 (or 26.25)~MeV/c ($\Delta p/p=0.07$
FWHM) were detected by a $127\:\mu\mbox{m}$ scintillator (1) before
traversing a $25\:\mu\mbox{m}$ stainless steel vacuum isolation
window.
The muons continued to lose energy while passing through the cryostat
70~K heat shield to eventually stop either in the gold target support
foil (2) or in the $\approx 400-800 \,\mu\mbox{m}$ thick solid
hydrogen target where they finally formed muonic atoms.
The hydrogen target frozen on foil (2), which was placed
perpendicularly to the muon beam axis, was called the upstream target
(US), and was made of pure protium or of protium with a small
admixture of deuterium (or tritium), depending on the experiment (see
Figs.~\ref{fig:schemb} and \ref{fig:schema}).

In the case of a pure protium US target, a thin neon layer was
additionally deposited on top, as represented in
Fig.~\ref{fig:schemb}.
In other cases, when deuterium or tritium were present in the US
target, we used an additional downstream target (DS) frozen on a
second gold foil placed parallel to the first foil but 17.9~mm further
along the beam axis ((3) in Fig.~\ref{fig:layout}).
Such an arrangement is presented in Fig.~\ref{fig:schema}, where a
thin layer of neon is shown sandwiched between a layer of pure protium
and the DS gold foil, and was used for the TOF studies of the
Ramsauer--Townsend effect.
A low muon beam momentum was chosen to minimize the number of $\mu$
stops in the DS protium layer.

Neon was used to detect the scattered muonic atoms which left the
hydrogen layer and subsequently transferred the muon to the neon.
The resulting emission of 207~keV x~rays from the 2p--1s $\mu$Ne
transition was observed by two $\sim$100~cm$^3$ germanium crystals (G1
and G2, Fig.~\ref{fig:layout}) with a time resolution of $10-12$~ns
(FWHM).
The G1 detector was used during the whole experiment, during both the
deuterium and the tritium measurements.
However, there were two physically different G2 detectors, one for
each of the deuterium and tritium measurements.
The plastic scintillators ((7) in Fig.~\ref{fig:layout}) were located
around the target to detect the muon decay electrons.
The NaI detector was used to study $\mu$CF in hydrogen and deuterium
mixtures~\cite{olinx99,mulha99}.  

\subsection{The method}
\label{sec:method}

\begin{table*}[t]
\begin{ruledtabular} 
\caption{Different measurements performed for the RT and
$\mathrm{p}\mu$ diffusion studies.  
DEm (Deuterium Emission) stands for a layer of
$1500\:\mbox{Torr}\cdot\ell (\mathrm{H}_2 + 0.05\% \mathrm{D}_2)$
covered with $500\:\mbox{Torr}\cdot\ell$ $\mathrm{H}_2$.  
TEm (Tritium Emission) --- $2000\:\mbox{Torr}\cdot\ell (\mathrm{H}_2 +
0.12\% \mathrm{T}_2)$.  
sTEm (Small Tritium Emission) --- $1000\:\mbox{Torr}\cdot\ell
(\mathrm{H}_2 + 0.12\% \mathrm{T}_2)$.
PP(Pure Protium) ---
$2000\:\mbox{Torr}\cdot\ell$ $\mathrm{H}_2$.  
sPP(Small Pure Protium)
--- $1000\:\mbox{Torr}\cdot\ell$ $\mathrm{H}_2$.  
GMU --- Good Muons: i.e., events when only one muon entered the
apparatus (no pileup).
Conversion factor (for hydrogen): $1\:\mbox{Torr}\cdot\ell$
corresponds to $3.4\:\mu\mbox{g}\cdot\mbox{cm}^{-2}$ for
$\mathrm{H}_2$.}
\label{tab:rt}
\begin{tabular}{llcccccc}
Label &Experimental&Beam & US Hydrogen & US Neon & DS Protium & DS Neon & GMU \\
      & Purpose    &MeV/c& $\mbox{Torr}\cdot\ell$ & $\mbox{Torr}\cdot\ell$ 
                   & $\mbox{Torr}\cdot\ell$ & $\mbox{Torr}\cdot\ell$ & $\times 10^6$\\ \hline
D1    & RT       & 26.70 & DEm         & ---  & --- & 100 & 326.9\\
D2    & RT       & 26.70 & DEm         & ---  & --- &  50 & 183.3\\
D3    & RT, diff & 26.70 & DEm         & ---  & 300 &  50 & 521.8\\
D4    & RT, diff\footnotemark[1]&26.70&DEm&---& 600 &  50 & 433.2\\
D5    & diff     & 26.70 & DEm         & 100  & --- & --- & 96.6\\
D6    & diff     & 26.70 & DEm         &  50  & --- & --- & 136.9\\
D7    & diff     & 26.70 & PP          & ---  & 300 &  50 & 149.4\\
T1    & RT       & 26.25 & TEm         & ---  & --- &  30 & 113.5\\
T2    & RT       & 26.25 & TEm         & ---  & --- &  50 & 174.2\\
T3    & RT, diff\footnotemark[1]&26.25&TEm&---& 350 &  50 & 405.3\\
T4    & RT, diff & 26.25 & sTEm        & ---  & 500 &  50 & 147.1\\
T5    & diff     & 26.25 & sPP         & 10   & --- & --- & 199.3\\
T6    & diff     & 26.25 & sPP         & 20   & --- & --- & 195.8\\
\end{tabular}
\footnotetext[1]{D4 and T3 are not useful for the $\mathrm{p}\mu$
diffusion analysis due to the strong overlap between RT and diffusion
parts of the time spectra.}
\end{ruledtabular} 
\end{table*}

\begin{figure}[b]
\includegraphics[angle=-90,width=0.24\textwidth]{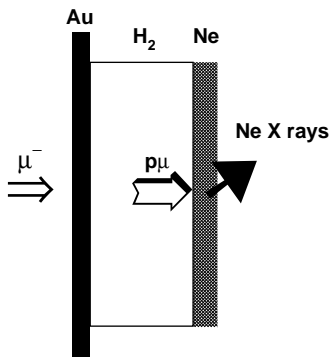}
\caption{Single target scheme.}
\label{fig:schemb}
\end{figure}

\begin{figure}[b]
\includegraphics[angle=-90,width=0.48\textwidth]{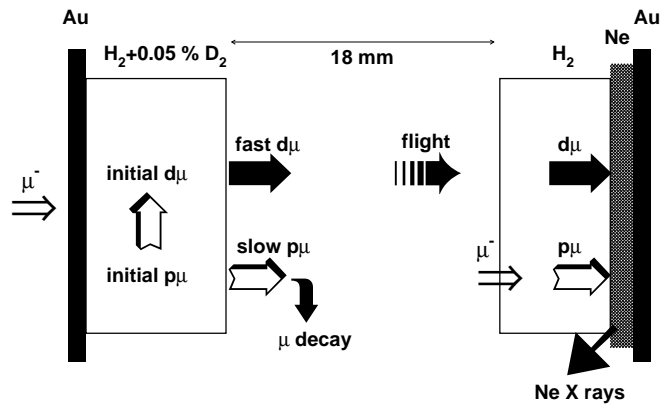}
\caption{Double target TOF scheme for $\mathrm{d}\mu$.  Fast and slow
$\mathrm{d}\mu$ atoms were emitted forward into the vacuum gap.
Although not shown in the drawing, muonic atoms were also emitted
toward the gold foils, which we refer to as backward emission.}
\label{fig:schema}                                             
\end{figure}

Following muon capture and $\mathrm{p}\mu$ formation, the
$\mathrm{p}\mu$ atoms slow down and diffuse in the hydrogen layer,
with some significant fraction of the $\mathrm{p}\mu$ atoms escaping
the layer.
Analyzing the emission yield and the time distribution of the escaped
$\mathrm{p}\mu$'s gives information about the scattering cross
sections.
The essential part of the analysis is the comparison of the
experimental yields and time distributions with the ones calculated by
Monte Carlo.
The measurements were performed in two different ways; either by using
the single pure protium target covered with a Ne layer
(Fig.~\ref{fig:schemb}) or by using the parallel target scheme
(Fig.~\ref{fig:schema}).

The first method (Fig.~\ref{fig:schemb}) proved better for studying
$\mathrm{p}\mu$ scattering for two reasons: (1) high stopping rate in
the US target assured high $\mathrm{p}\mu$ emission statistics, and
(2) the time spectrum was ``clean'' in that it did not contain the
overlapping RT part.
Table \ref{tab:rt} summarizes the different measurements performed for
the study of the RT effect and $\mathrm{p}\mu + \mathrm{p}$ scattering.

In the second case (Fig.~\ref{fig:schema}), the US target was composed
of protium with a small concentration of $\mathrm{D}_2$ (or
$\mathrm{T}_2$), which served as a source of energetic $\mathrm{d}\mu$
($\mathrm{t}\mu$) atoms emitted from the layer into the adjacent
vacuum with a mean energy of 3.5~eV (9~eV in the case of
$\mathrm{H}_2$/$\mathrm{T}_2$ mixture) as a result of the RT effect.
Due to the low deuterium (tritium) concentration, $\mathrm{p}\mu$'s
were predominantly formed as a result of muon stops in the US layer.
A small fraction of them survived their evolution in the US hydrogen (no
muon transfer to the heavier hydrogen isotope, no $\mathrm{pp}\mu$
formation, or muon decay) and left the solid layer after multiple
scattering.
However, they were very slow ($\varepsilon \approx$ meV) and could not 
reach the downstream target before the muon decayed.
In most cases the muon was transfered from $\mathrm{p}\mu$ to a
deuterium (tritium) atom.
At formation, the muonic deuterium and tritium atoms had a relatively
high kinetic energy, about 45~eV, which they subsequently lost in
elastic collisions, mainly with protium, until the energy reached the
range of the Ramsauer--Townsend (RT) minimum in the scattering cross
section $\sigma(\mu\mathrm{d} +\mathrm{H}_2$).
Then the mean distance between collisions increased and the hydrogen
layer became effectively transparent for the $\mathrm{d}\mu$
($\mathrm{t}\mu$) atoms which were then easily emitted from the solid
into the adjacent vacuum.
Such energetic emitted muonic atoms traveled through vacuum toward the
downstream hydrogen layer (DS) and a fraction passed, after possible
interactions in hydrogen, to the Ne layer and produced x~rays as a
result of the muon transfer to neon and subsequent muonic neon
deexcitation.
Such $\mathrm{d}\mu$ ($\mathrm{t}\mu$) atoms gave a characteristic
peak in the TOF spectrum at $\approx1\:\mu\mathrm{s}$ --- a time
determined by the distance between the foils and the position in
energy of the Ramsauer--Townsend minimum.

Clearly, not all muons were stopped in the US target and thus those
which reached the second foil created $\mathrm{p}\mu$'s in the DS
protium.
The $\mathrm{p}\mu$'s diffused to neon giving a contribution to the
time spectrum at early times.
Thus the resulting Ne time spectrum contained two relatively distinct
components, one of them connected with the RT effect in
$\mathrm{d}\mu$ ($\mathrm{t}\mu$) scattering and the other with the
diffusion of $\mathrm{p}\mu$ atoms in the solid hydrogen.
Despite the overlap of the two effects in the time spectra one should
note there is an important advantage of such an experiment.
Since the kinetic energy of the $\mathrm{d}\mu$ ($\mathrm{t}\mu$) is
relatively high, the delayed RT peak is not sensitive to the state of
the target material and can be well described using either ``gas'' or
``solid'' scattering cross sections (see~\cite{mulha99,mulha01}).
Due to this, the RT peak can be used as a reference in the analysis of
the diffusion part where the effects of the solid state can be found.

\subsection{Monte Carlo simulations}
\label{sec:monte-carlo-simul}

The Monte Carlo code FOW~\cite{wozni96} was used in the planning
stages of the experiment as well as for the analysis to simulate all
physical processes occurring after muons pass through the entrance
window of the apparatus.
Muon stopping distributions along the beam axis in the different
apparatus components, especially in the hydrogen layers of the target, 
have been taken from a special set of measurements~\cite{mulha99} and
from another Monte Carlo calculation~\cite{jacot97b} and used as an
input to FOW\@.
The FOW code gives the possibility of a full three--dimensional
description of the target geometry.

The muonic processes considered are as follows: 
\begin{itemize}
  \item[(i)] elastic scattering: 
    $\mathrm{p}\mu+\mathrm{p}$, 
    $\mathrm{p}\mu+\mathrm{d}$,
    $\mathrm{d}\mu+\mathrm{p}$,
    $\mathrm{p}\mu+\mathrm{t}$,
    $\mathrm{t}\mu+\mathrm{p}$,
    $\mathrm{d}\mu+\mathrm{d}$,
    $\mathrm{t}\mu+\mathrm{t}$.
  \item[(ii)] spin--flip: 
    $\mathrm{p}\mu(\mathrm{F}) \rightarrow \mathrm{p}\mu(\mathrm{F}')$, 
    $\mathrm{d}\mu(\mathrm{F}) \rightarrow \mathrm{d}\mu(\mathrm{F}')$,
    $\mathrm{t}\mu(\mathrm{F}) \rightarrow \mathrm{t}\mu(\mathrm{F}')$.
  \item[(iii)] charge transfer: 
    $\mathrm{p}\mu \rightarrow \mathrm{d}\mu$,
    $\mathrm{p}\mu \rightarrow \mathrm{t}\mu$.
  \item[(iv)] molecular formation: 
    $\mathrm{p}\mu+\mathrm{p} \rightarrow \mathrm{pp}\mu$, 
    $\mathrm{d}\mu+\mathrm{d} \rightarrow \mathrm{dd}\mu$, 
    $\mathrm{t}\mu+\mathrm{t} \rightarrow \mathrm{tt}\mu$, 
    $\mathrm{d}\mu+\mathrm{p} \rightarrow \mathrm{pd}\mu$,
    $\mathrm{t}\mu+\mathrm{p} \rightarrow \mathrm{pt}\mu$.
\end{itemize}

\begin{table}[t]
\caption{Characteristics of the muonic processes in different upstream
targets (see column 4 of Table~\ref{tab:rt} for target details) as
calculated by the Monte Carlo.  
Muon stops in the targets are given in \% of muons entering the
apparatus (the Monte Carlo analog of GMU)\@.
The emission yield, molecular formation, backward escape, and muon
decay are given in \% per muon stopped.
Simulations were performed using the ``solid'' cross sections.}
\label{tab:targUS}
\begin{ruledtabular}
\begin{tabular}{lrrrrr}
upstream targets                 & DEm  & TEm   & sTEm  & sPP   & PP   \\ \hline 
$\mu$--stops                      & 58.6  & 47.4  & 32.3  & 32.3  & 58.6 \\
$\mathrm{pp}\mu$ formation       & 34.1  & 24.3  &  4.8 & 77.9  & 83.8 \\
$\mathrm{pd}\mu$ formation       & 39.7  & ---   & ---   & ---   & ---  \\
$\mathrm{pt}\mu$ formation       & ---   & 48.5  & 35.8  & ---   & ---  \\
muon decay                       &  7.1 &  6.1 &  2.8 &  9.5 & 10.2 \\
backward escape                  & 13.0  & 15.2  & 31.0  &  8.1 & 4.1 \\
forward $\mathrm{p}\mu$ emission &  1.8 &  0.4 &  0.4 &  4.5 & 1.9 \\
forward $\mathrm{d}\mu$ emission &  4.3 & ---   & ---   & ---   & ---  \\
forward $\mathrm{t}\mu$ emission & ---   &  5.5 & 25.2  & ---   & ---  \\
\end{tabular}
\end{ruledtabular}
\end{table}

\begin{figure}[b]
\includegraphics[width=0.48\textwidth]{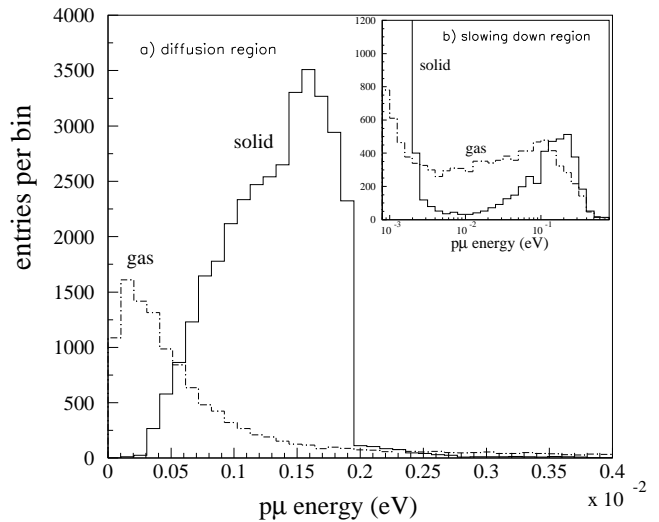}
\caption{Calculated low energy spectrum (a) of emitted $\mathrm{p}\mu$
atoms for the PP target.  
The solid line shows the result using the ``solid'' cross sections,
whereas the dot--dashed line shows the spectrum using the ``gas''
cross sections.
The same number of incident muons ($2\cdot 10^6$) is used for both
cases.  
Top--right picture (b) shows the slowing down energy spectrum in a
log--binned scale.}
\label{fig:energy-pp-compl}
\end{figure}

Energy--dependent values of the total and differential cross sections
for the elastic scattering of muonic atoms, spin--flip, and charge
transfer transitions were used in the calculations.
For small collision energies (usually below 0.1~eV) the ``solid''
double differential cross sections~\cite{adamc97} were used for the
elastic scattering and spin--flip interactions.
At higher energies, where the solid state effects become negligible,
the total and single differential ``gas'' cross sections from
Refs.~\cite{bracc89b,melez92,chicc92,melez98}, corrected for molecular
effects via the Sachs--Teller model, were applied.
Using ``gas'' cross sections at higher energies saved computer time
without incurring any loss in accuracy.
The energy--dependent $\mathrm{pd}\mu$, $\mathrm{pt}\mu$ and
$\mathrm{dd}\mu$ (resonant and nonresonant) formation rates were taken
from the Faifman calculations~\cite{faifm89b,faifm88,ponom76}.
The $\mathrm{pp}\mu$ and $\mathrm{tt}\mu$ formation rates were
considered as energy independent ($\lambda_{\mathrm{pp}\mu} =
3.21\:\mu\mbox{s}^{-1}$~\cite{mulha96}, $\lambda_{{\mathrm tt}\mu} =
1.80\:\mu\mbox{s}^{-1}$~\cite{breun87c}).
Thermal motion of the target molecules was also taken into account.
Tables~\ref{tab:targUS} and~\ref{tab:targDS} show the main
characteristics calculated for different upstream and downstream
targets used for $\mathrm{p}\mu$ emission study.
The values are based on at least $10^6$~simulated muons, so the
statistical uncertainty is negligible.

\begin{table}[t]
\caption{Characteristics of the muonic processes in downstream protium
targets (see column 6 of Table~\ref{tab:rt} for target name
references).  
Muon stops, $\mathrm{p}\mu$ emission and $\mathrm{d}\mu$
($\mathrm{t}\mu$) transmission are given in \% of muons passing the
entrance window of the apparatus.
Simulations are performed using the ``solid'' cross sections.}
\label{tab:targDS}
\begin{ruledtabular}
\begin{tabular}{lrrrr}
 downstream targets              & 600  & 500  & 350  & 300  \\
(associated US target)           & DEm & sTEm & TEm  & DEm \\ \hline
$\mu$--stops                      & 7.8 & 9.3 & 2.6 & 4.4 \\
forward $\mathrm{p}\mu$ emission & 0.6 & 0.9  & 0.4  & 0.7  \\
$\mathrm{d}\mu$ transmission     & 0.7 & ---  & ---  & 0.9 \\
$\mathrm{t}\mu$ transmission     & ---  & 1.4 & 0.9 & ---  \\
\end{tabular}
\end{ruledtabular}
\end{table}

\begin{figure}[b]
\includegraphics[width=0.48\textwidth]{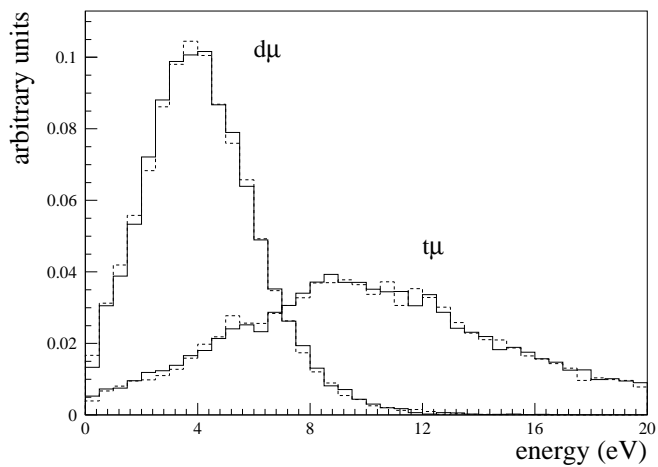}
\caption{Energy spectra of $\mathrm{d}\mu$ and $\mathrm{t}\mu$ atoms
emitted from DEm and TEm targets, respectively, after flying the
US--DS distance and before entering the DS target. Solid lines: MC
with the ``solid'' cross sections, dashed lines: MC with the ``gas''
cross sections.}
\label{fig:pl74}
\end{figure}

Figure~\ref{fig:energy-pp-compl} shows the calculated energy spectrum
of $\mathrm{p}\mu$ atoms emitted from the PP target.
A strong solid--state effect is evident both in yield intensity and
spectral shapes when comparing the calculations which include or
disregard the solid hydrogen structure (solid and dot--dashed lines,
respectively).
The details of the spectral tail above the Bragg cutoff limit are
shown on a log--binned scale in the insert.
The kinetic energy spectra for $\mathrm{d}\mu$ and $\mathrm{t}\mu$
atoms emitted from DEm and TEm targets, respectively, are shown in
Fig.~\ref{fig:pl74} for the ``solid'' (solid line) and ``gas'' (dashed
line) cross sections.
The lack of solid state effects is not surprising given the high
energy of the atoms involved.
The muonic atom energy applies to atoms which have traveled the
17.9~mm distance between the US and DS layers.

\begin{figure}[t]
\includegraphics[width=0.48\textwidth]{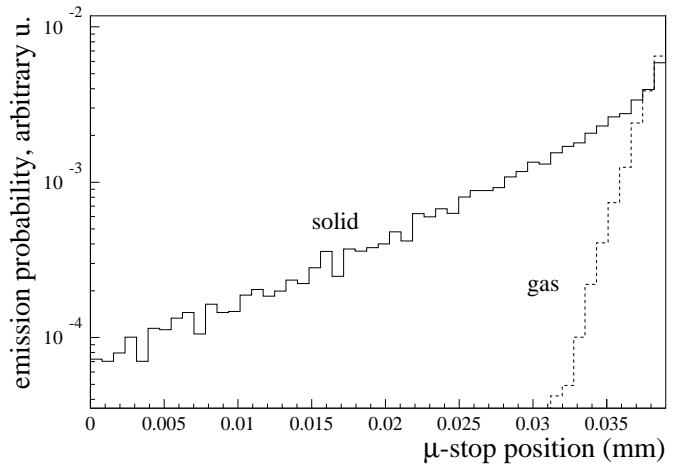}
\caption{The probability dependence of $\mathrm{p}\mu$ emission from
the PP target (for a beam momentum $\mathrm{p}=26.25$~MeV/c) versus
the initial $\mathrm{p}\mu$ position in the target.  
The simulation was performed using both the ``solid'' cross sections
(solid line) and ``gas'' cross sections (dashed line) for the same
number of incident muons.}
\label{fig:position}
\end{figure}

Figure~\ref{fig:position} shows the probability of $\mathrm{p}\mu$
emission from a hydrogen target as a function of the initial
$\mathrm{p}\mu$ formation position following the muon stop.
The example is for the PP target and for the muon stopping distribution
for the beam momentum 26.25~MeV/c.
The difference between the results of the ``solid'' cross sections
(solid line) and ``gas'' cross sections (dashed line) illustrates the
strong increase of the mean free path in the final stage of the
$\mathrm{p}\mu$ slowing down when the solid state effects are
considered.
One can see that the volume from which emitted $\mathrm{p}\mu$ atoms
can originate is much more extended in the solid case.

\section{Measurements}
\label{sec:measurements}

\begin{figure}[t]
\includegraphics[angle=90,width=0.48\textwidth]{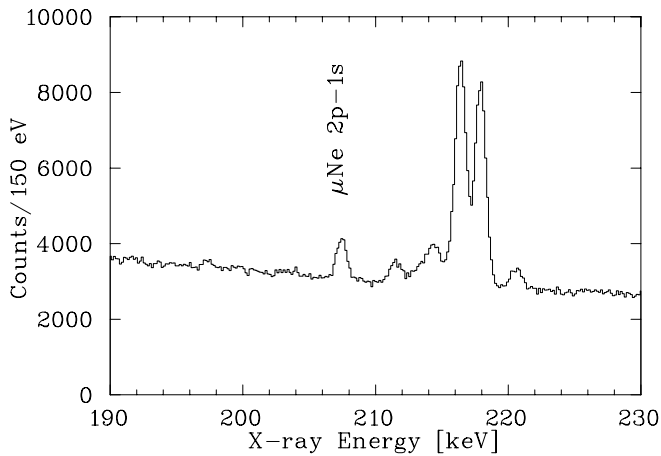}
\caption{x--ray energy spectrum for the T3 measurement.  The $\mu$Ne
2p--1s is located at 207~keV, whereas the bigger peaks around
$215-220$~keV are the $\mu$Au 6--5 lines.}
\label{fig:xrayene}
\end{figure}

Muonic atom scattering in hydrogen was measured via the x--ray time
spectra of $\mu$Ne 2p--1s at 207~keV (see Fig.~\ref{fig:xrayene}).
The time spectra events were selected within the energy window
$205.6-208.3$~keV\@.
The background time spectra under the 207~keV peak were created using
time spectra from two neighboring energy windows, namely a left
background at $203.3-205.6$~keV and a right background at
$208.3-210.2$~keV\@.
Two different background evaluation procedures were used.
In the first, the left and right spectra were added, and then
normalized by the energy window widths and the resulting spectrum
subtracted from the time spectrum of the $\mu$Ne 2p--1s peak.
The second relied on a multi--parameter fit of the summed left and
right backgrounds using two exponential functions (with lifetimes for
muons in gold and neon) and the background predicted from the fit
function was subtracted from the $\mu$Ne 2p--1s peak.
Since the background accounts for $70-85$\% of the total statistics, its
removal plays an important role, especially for data at early times
where the muon prompt capture in the neon layer and the gold foils bring
a strong contribution.
Another data cleaning method resulting in better signal to background
was the requirement that the muon decay electron be seen {\em after}
the $\mu$Ne 2p--1s x~ray, starting from a given time delay.
Those delayed electrons were detected by the scintillators (see
Fig.~\ref{fig:layout}) during a time interval $0.2-5.2 \,\mu\mbox{s}$
after the $\mu$Ne signal.
This method, called the ``del$_e$'' criterion, suppressed the
background by a factor of about 300. However, useful statistics were
reduced by about a factor ten.

\begin{table*}[t]
\caption{Experiments for the $\mathrm{p}\mu+\mathrm{p}$ scattering
study.  
When the hydrogen layer had the same thickness, the different
measurements were summed to give more statistics.
The last column indicates the signals present in the time spectrum.}
\label{tab:useful}
\begin{ruledtabular}
\begin{tabular}{clrcr}
No. & Label & Hydrogen & GMU & Time Spectrum \\
 \#      & & [$\mbox{Torr}\cdot\ell$] & $\times 10^6$ &  \\ \hline  
1      & D3+D7 & 300 DS & 671  & $\mathrm{p}\mu$ + delayed $\mathrm{d}\mu$ from DEm \\
2      & T4    & 500 DS & 147  & $\mathrm{p}\mu$ + delayed $\mathrm{t}\mu$ from sTEm \\
3      & T5+T6 & 1000 US & 395 & $\mathrm{p}\mu$  clean spectrum \\
4      & D5+D6 & 2000 US & 233 & $\mathrm{p}\mu$ + $\mathrm{d}\mu$ diffusion from DEm \\
\end{tabular}
\end{ruledtabular}
\end{table*}

\subsection{Combined measurements with H/D and H/T targets}
\label{sec:comb-meas-with}

\begin{figure}[b]
\includegraphics[width=0.48\textwidth]{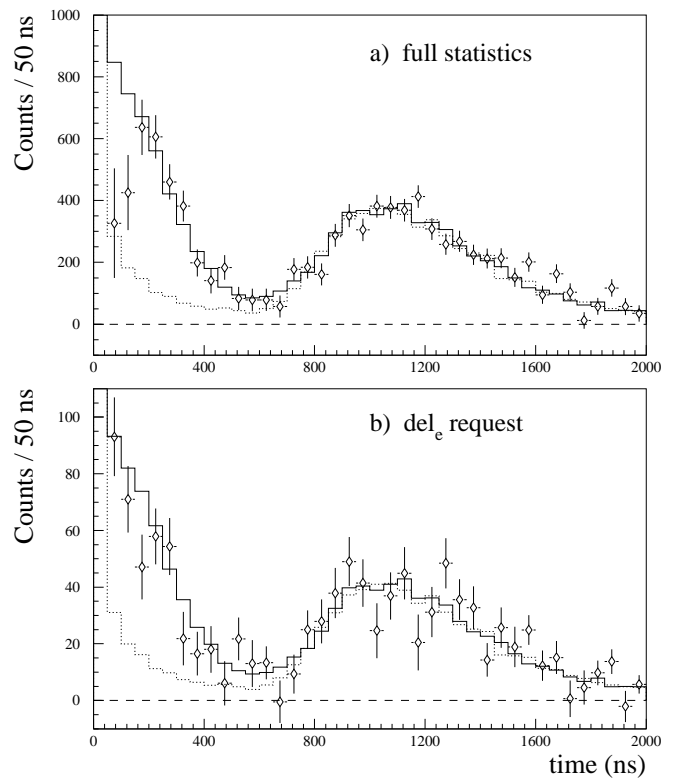}
\caption{Experimental time--of--flight spectra (points with error
bars) for experiment \#1 (see Table~\ref{tab:useful}) for cases: a)
full statistics, b) del$_e$ criteria.  
The solid line represents the Monte Carlo simulation based on the
scattering cross sections when solid effects were taken into account,
the dotted line is for the ``gas'' cross sections.}
\label{fig:rt300}
\end{figure}

For the $\mathrm{p}\mu$ scattering analysis only the few experiments
from Table~\ref{tab:rt} in which the $\mathrm{p}\mu$ diffusion time
spectrum have been seen are useful.
Similar runs were summed whenever possible and Table~\ref{tab:useful}
give the details.
The other measurements were nevertheless necessary for the
determination of detection efficiencies.

A typical TOF measurement (\#1, Table~\ref{tab:useful}), where both
$\mathrm{d}\mu$'s from the US layer and $\mathrm{p}\mu$'s from the DS
protium layer were detected when they reach the DS neon, is presented
in Fig.~\ref{fig:rt300} (points with error bars).
The a) and b) graphs show the time spectra for the full statistics and
for the events where the del$_e$ criterion has been applied for
background suppression, respectively.
That measurement was performed with a DEm upstream target (with
$0.05\%$ $\mathrm{D}_2$) and a $300\:\mbox{Torr}\cdot\ell$
($\mathrm{H}_2$) downstream layer.
The events occurring at early times ($t < 600$~ns) are due to
$\mathrm{p}\mu$ formed directly in the DS hydrogen which then diffuse
to the neon layer.
The peak in the TOF spectrum corresponds to the delayed
$\mathrm{d}\mu$ atoms which travel the distance between the two foils
and are not stopped in the DS hydrogen due to the RT effect.
Also plotted are the simulations using the ``solid'' scattering cross
sections as well as the result of the calculation when one neglects
the solid state effects and uses only the ``gas'' cross sections.

\begin{figure}[b]
\includegraphics[width=0.48\textwidth]{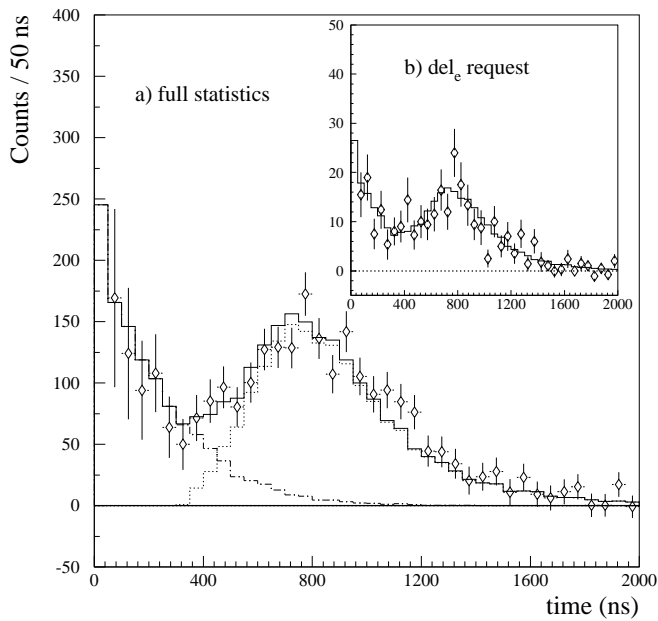}
\caption{Experimental time--of--flight spectra (points with error
bars) for experiment \#2 (see Table~\ref{tab:useful}) for a) full
statistics and b) del$_e$ statistics.  
The solid line is the MC simulation based on the scattering cross
sections when solid state effects were taken into account. 
Dot--dashed and dotted lines show the $\mathrm{p}\mu$ and
$\mathrm{t}\mu$ contributions, respectively.}
\label{fig:rt500}
\end{figure}

Another example of a similar TOF measurement (with a hydrogen/tritium
mixture in the target upstream and $500\:\mbox{Torr}\cdot\ell$ protium
covering the downstream Ne, \#2) is shown in Fig.~\ref{fig:rt500}
for the full statistics and del$_e$ requirement cases.
A relatively high muon stopping fraction in the downstream target
(because the US target was only $1000\:\mbox{Torr}\cdot\ell$ and 9.3\%
of the muons stopped in the DS target) gives good statistics for the
$\mathrm{p}\mu$ part in the time spectrum.
The delayed peak from $\mathrm{t}\mu$ transfer events lies earlier in
time than the corresponding deuterium case because of the higher
$\mathrm{t}\mu$ energy (see Fig.~\ref{fig:pl74}), and hence the
overlap of both diffusion and RT spectra parts is fairly strong.
The ``solid'' cross section MC spectrum is also presented in the
figure (solid line histogram).
Dot--dashed and dotted lines show the predicted contributions from
$\mathrm{p}\mu$ and $\mathrm{t}\mu$, respectively.

\begin{figure}[b]
\includegraphics[angle=90,width=0.48\textwidth]{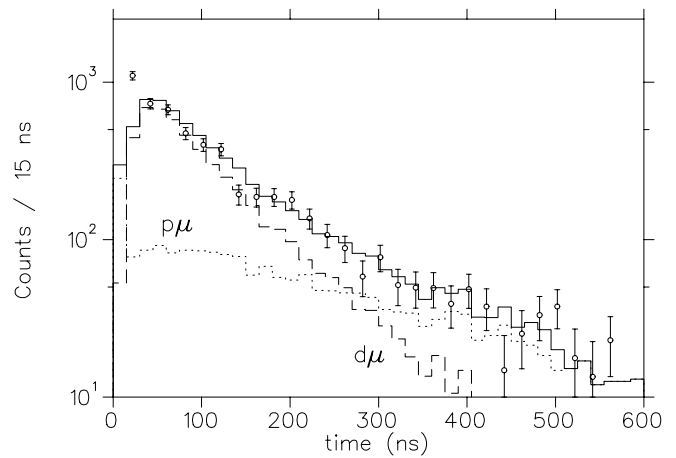}
\caption{Experimental time spectrum (points with error bars) of
$\mathrm{p}\mu$ and $\mathrm{d}\mu$ emitted together from DEm upstream
target (experiment \#4, Table~\ref{tab:useful}).  
Monte Carlo simulation is shown as the solid line histogram.
Dotted and dashed curves show the contributions from $\mathrm{p}\mu$ and
$\mathrm{d}\mu$, respectively.
Calculations based on ``solid'' cross sections.}
\label{fig:x2000}
\end{figure}

Figure~\ref{fig:x2000} shows the time spectrum of $\mathrm{p}\mu$ and
$\mathrm{d}\mu$ atoms emitted together from a DEm target (\#4,
Table~\ref{tab:useful}).
This case was analyzed using the different emission time dependences
since the $\mathrm{d}\mu$ part of the time spectrum decreases much
faster than the $\mathrm{p}\mu$ emission spectrum; the mean diffusion
time of $\mathrm{d}\mu$ in the $2000\:\mbox{Torr}\cdot\ell$ target is
$\sim100$~ns, much less than for the $\mathrm{p}\mu$ emission
($\sim300$~ns).
The $\mathrm{d}\mu$ contribution can be removed using the MC simulations
since $\mathrm{d}\mu$ emission is independent of solid state effects as
one can see from Fig.~\ref{fig:rt300}.
The subtraction leaves a clean $\mathrm{p}\mu$ spectrum which can be
compared to the MC including solid state effects.

Normalization of the simulated time spectra was done using the RT peak
which is equally well described by both the ``solid'' and ``gas''
approaches.
The normalization factor $n=Y(exp)/Y(MC)$, where the yields $Y(exp)$
and $Y(MC)$ were the total counts in the time interval $500-2500$~ns,
was calculated from all eight measurements given in Table~\ref{tab:rt}
which deal with the RT effect.
The mean weighted value for the germanium detector G1 (which was the
same in all experimental runs) was $n=(5.8\pm 0.3)\times 10^{-4}$.
This normalization factor, which is effectively the detector
efficiency, applies to Ne x~rays detected from the DS neon layer.
For experiments \#3 and~4 (see Table~\ref{tab:useful}) where x~rays
were detected from neon on the US target (shifted by 17.9~mm compared
to the DS target) the efficiency was $\approx40$\% higher (established
by comparing the total counts in the prompt Ne x--ray peaks in
experiments D1 and D5) and the value for detector G1 was $n=(8.1\pm
1.0)\times 10^{-4}$.
No efficiency was determined for the G2 germanium detector, because it
was changed between the deuterium and tritium measurements.

\subsection{Emission of $\mathrm{p}\mu$ atoms from a layer of solid hydrogen}
\label{sec:emission-rmpmu-atoms}

Although $\mathrm{p}\mu$ atom emission was a parasitic process in the
RT experiments, those runs were still useful for the analysis of
$\mathrm{p}\mu$ scattering in solid hydrogen.
In particular, the profound disagreement between experimental data and
theory using ``gas'' cross sections and the relative correctness of
the ``solid'' cross sections can be seen.

The observation of enhanced $\mathrm{p}\mu$ emission from solid hydrogen
stimulated additional measurements specifically intended to study the
phenomenon more precisely with higher statistics.
A target similar to the one shown in Fig.~\ref{fig:schemb} was made
from $1000\:\mbox{Torr}\cdot\ell$ pure protium covered with a thin
($10-20 \,\mbox{Torr}\cdot\ell$) neon layer.
The resulting time spectrum of $\mu$Ne 2p--1s x~rays described the
diffusion of $\mathrm{p}\mu$ atoms in hydrogen from the moment of the
muon stop to the moment of emission.

\begin{figure}[t]
\includegraphics[width=0.48\textwidth]{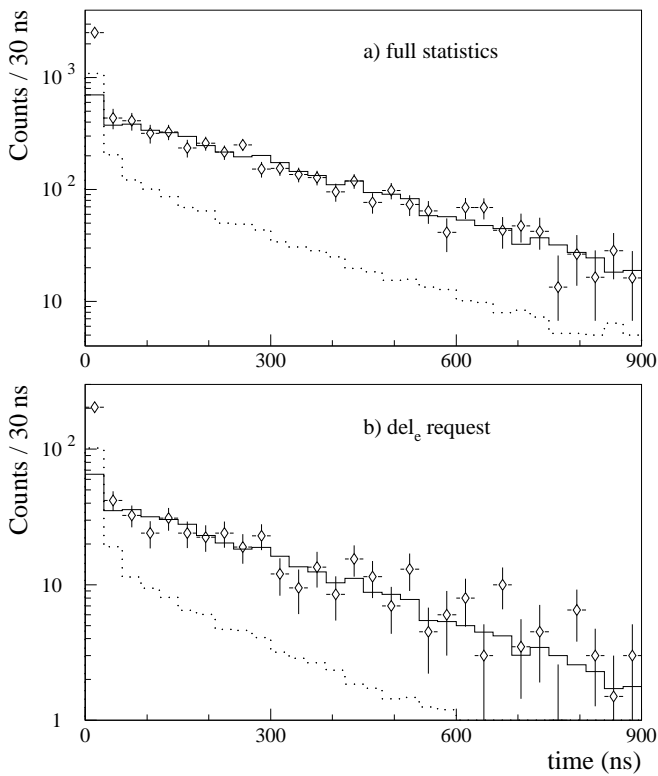}
\caption{Experimental (points with error bars) and MC (lines) time
spectra of $\mathrm{p}\mu$ emitted from the
$1000\:\mbox{Torr}\cdot\ell$ $\mathrm{H}_2$ layer (\#3,
Table~\ref{tab:useful}) for the full experimental statistics, a), and
del$_e$ criteria, b).
Solid line: calculation with the ``solid'' cross sections; dotted
line: calculation for the ``gas''cross sections.
Calculations with ``solid'' cross sections are normalized to the
experimental data according to conclusions from the TOF measurements.}
\label{fig:difa}
\end{figure}

The measured time spectra (\#3, Table~\ref{tab:useful}) of $\mu$Ne
2p--1s x~rays are shown in Fig.~\ref{fig:difa} for the full statistics
and del$_e$ criteria.
MC spectra also shown on the figures describe well the experimental
data when the ``solid'' cross sections are used (solid lines).
Normalization to the experimental data is based on the conclusions
from the TOF measurements (see Sec.~\ref{sec:comb-meas-with}).
The calculation with the ``gas'' cross sections (dotted lines) gives a
suppressed yield although the lifetimes representing the diffusion are
not dramatically different.

\section{Analysis and discussion}
\label{sec:analysis-discussion}

As is seen from the comparison between the experimental results and
the simulations performed using both the ``solid'' and ``gas'' cross
sections presented on Fig.~\ref{fig:rt300}, both types of cross
section describe the RT part of the spectrum equally well.
That agreement is due to the fact that solid state effects are
negligible for the energies of the $\mathrm{d}\mu$ atoms which are
responsible for those spectra.
A similar conclusion can be drawn from the measurements with
$\mathrm{t}\mu$.

In contrast, there is a big difference in yields as well as in time
dependence between the calculated $\mathrm{p}\mu$ diffusion spectra
and measurements where an agreement with the experiment is obtained
only for the ``solid'' cross sections; the discrepancy seen in
Fig.~\ref{fig:rt300}~a) for first points from times $0-150$~ns has an
artificial source and will be explained later in this section.
The ``gas'' cross sections predict a total yield of emitted
$\mathrm{p}\mu$'s two times smaller than required for a correct
description of the experimental results.
If one excludes the first channel (Fig.~\ref{fig:rt300}), which
contains events connected with fast $\mathrm{p}\mu$ atoms from the
slowing down stage, the ``gas'' approach gives three times less
emission.

The good agreement between experiment and calculation based on the
``solid'' cross sections is also visible in Fig.~\ref{fig:rt500} where
the emission spectrum of $\mathrm{p}\mu$'s from the DS target and the
transmission spectrum of the delayed $\mathrm{t}\mu$'s from the sTEm
US target are presented.
The $\mathrm{p}\mu$ emission is enhanced 2.3~times for the total
emission and and 3.8~times for emission at times $t > 30$~ns.
Accurate results, especially from the point of view of the emitted
$\mathrm{p}\mu$ diffusion time analysis, are obtained from the
experiment with the sPP target (Fig.~\ref{fig:difa}).
In that case there were no ambiguities between signals from diffusion
and those from RT events, which was a problem present in the
measurements with the combined H/D and H/T targets.
However such combined target measurements were necessary because they gave
the RT peak which was used as the reference for the yield
normalization of the MC spectra.
The time distribution of the emitted $\mathrm{p}\mu$ atoms has some
nontrivial behavior and depends not only on the cross sections but
also on the target thickness, the muon stop distribution in the
target, and the initial energy of the $\mathrm{p}\mu$ atoms.
In the limit of long times and for a given $\mu$--stop distribution,
the $\mathrm{p}\mu$ emission time distribution can be well modeled by
a one--exponential approximation.
In that limit, the constant factor in the exponent represents the mean
time needed for the equilibrated $\mathrm{p}\mu$ atoms to reach the
layer boundary (we call that parameter the diffusion time, $\tau_d$).
Such a $\tau_d$ depends on the target thickness but should reach an
asymptotic value for high thickness simply because the $\mu$--stop
distribution is effectively exponential decreasing in the thickness
(c.f. Fig~1 of Ref~\cite{wozni96}).
When convoluted with the escape probability of
Fig.~\ref{fig:position}, and the associated escape time from any given
depth, this exponential behaviour of the stopping yields a constant
emission time.
In light of the remarks presented above, we chose a unique time region
$t > 200\:\mathrm{ns}$ for the analysis of the four experimental time
spectra (see Table~\ref{tab:useful}) to determine the diffusion time,
$\tau_d$ and compare it with the MC simulation.

Such a choice had the additional advantage that it avoided problems
with the early parts of the time spectrum.
The problem is clearly visible in Fig.~\ref{fig:rt300}~a) for times
less than 200~ns in a~measurement made with a thick US layer plus an
$\mathrm{H}_2$ DS layer.
The early time signal was only a few percent of the total counts so the
background subtraction for those time regions was significant, as can
be seen from the resulting uncertainties, and hence any small
irregularities in the thresholds and in the detection of secondary
muons as well as in the background estimation itself can result in an
poor estimation of the background.
It also exists, but is less important, in measurements with an
$\mathrm{H}_2$ US layer, where the percentage of incident muons
stopped US was significantly higher than in the downstream layer.
It is worth noting that for the spectra cleaned with the del$_e$
condition, and hence, where problems with background vanish, the
theory (via Monte Carlo simulations) agrees with the early time
region, but a more sophisticated analysis was not possible because of
poor statistics and the complicated form of the time spectra.

\begin{table}[t]
\caption{Diffusion time of $\mathrm{p}\mu$ atoms, $\tau_d$ in ns, in
solid hydrogen layers of different thicknesses (\#'s defined in
Table~\ref{tab:useful}).}
\label{tab:tdif}
\begin{ruledtabular}
\begin{tabular}{cccccc}
No.& Target & \multicolumn{2}{c}{Experiment} &
\multicolumn{2}{c}{MC\footnotemark[1]} \\
\cline{3-6} 
\#  & $\mbox{Torr}\cdot\ell$ & full stat. & del$_e$ &
 ``solid'' & ``gas'' \\ \hline
1 & 300 & 161(32) & 183(35) & 141(4) & 257(26) \\
2 & 500 & 199(40) & 360(165) & 222(7) & 234(14) \\
3 & 1000 & 269(13) & 279(44) & 258(3) & 237(8) \\
4 & 2000 & 279(26) & 253(42) & 254(8) & 230(24) \\
\end{tabular}
\end{ruledtabular}
\footnotetext[1]{statistical error is used for the fit}
\end{table}

\begin{figure}[t]
\includegraphics[angle=90,width=0.48\textwidth]{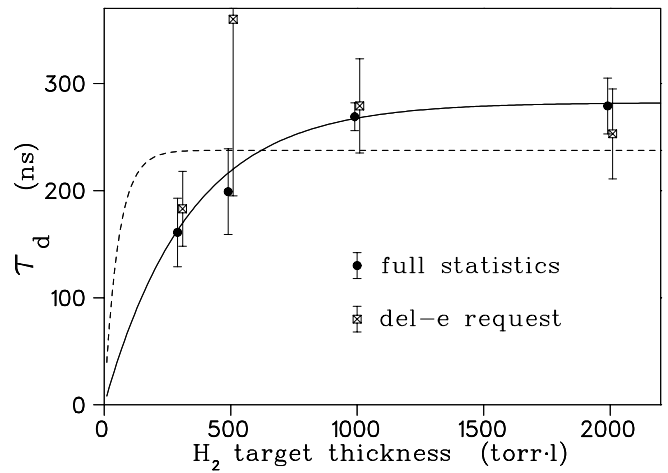}
\caption{The dependence of the diffusion time $\tau_d$ on
$\mathrm{H}_2$ layer thickness.  
Solid and dashed lines were calculations using ``solid'' and ``gas''
cross sections, respectively.}
\label{fig:tau}
\end{figure}

The measured and calculated values of $\tau_d$, fitted using a
single--exponential distribution, are given in Table~\ref{tab:tdif}
for both data treatments (i.e., with and without del$_e$).
Good agreement between the experimental values of $\tau_d$ and the
calculations using the ``solid'' cross sections is observed for each
experiment.
The results are also shown in Fig.~\ref{fig:tau} where the points with
error bars represent the experimental values of $\tau_d$ from
Table~\ref{tab:tdif}. 
The lines are the results from the Monte Carlo calculations using the
``solid'' and ``gas'' cross sections (solid and dotted lines,
respectively).
The MC results from Table~\ref{tab:tdif} were obtained strictly for
the given experimental conditions whereas the calculations represented
by the continuous lines in Fig.~\ref{fig:tau} were made assuming pure
protium layers of increasing thickness and using the same muon
stopping distribution (beam momentum of 26.25~MeV/c) in order to show
the smooth dependence of $\tau_d$ on the target thickness.

The analysis of the emission yields of the slow $\mathrm{p}\mu$ atoms
supports also the use of the ``solid'' cross sections. 
This is seen by comparing the experimental results with the
simulations when the ``solid'' and ``gas'' cross sections were used.
The results for the G1 detector are presented in Table~\ref{tab:yields}.
The comparison was performed for the time interval $200-600$~ns,
characteristic for the diffusion process, using the efficiencies
established from the RT time domain (see
Section~\ref{sec:comb-meas-with}).
The agreement between the experiments and ``solid'' cross sections is
excellent.

\begin{table}[b]
\caption{Comparison of the calculated and measured (full statistics)
$\mathrm{p}\mu$ atom emission yields (in \% per GMU) from the
different solid hydrogen layers for the time interval $200-600$~ns.
The second column represents the total hydrogen thickness in both US
and DS\@.}
\label{tab:yields}
\begin{ruledtabular}
\begin{tabular}{ccccc}
No. & Hydrogen & Experiment & \multicolumn{2}{c}{MC\footnotemark[1]}
\\
\cline{3-5} \# & Thickness &&``solid'' & ``gas''\\
& ($\mbox{Torr}\cdot\ell$) & & & \\ \hline
1 & 2000+300 & $0.21(2)$ & $0.225(5)$ & $0.059(2)$ \\
2 & 1000+500 & $0.35(6)$ & $0.345(6)$ & $0.067(3)$ \\
3 & 1000 & $0.52(7)$ & $0.521(7)$ & $0.313(6)$ \\
4 & 2000 & $0.38(6)$ & $0.398(6)$ & $0.074(3)$ \\
\end{tabular}
\end{ruledtabular}
\footnotetext[1]{Only the statistical error is given.}
\end{table}

Despite the general agreement between the experimental data and the
theoretical description of $\mathrm{p}\mu$ diffusion in fully modeled
solid hydrogen, the question of the sensitivity of the calculated
diffusion time and emission yield on the cross sections is important.
The influence of an inaccuracy in the $\mathrm{p}\mu +\mathrm{D}_2$
($\mathrm{T}_2$) and $\mathrm{d}\mu (\mathrm{t}\mu) + \mathrm{X}$ (X
is any hydrogen isotope) cross sections is negligible in the
$\mathrm{p}\mu$ diffusion study because of the very small
concentration of deuterium (tritium) admixtures in the targets.
The question of $\mathrm{d}\mu + \mathrm{H}_2$, $\mathrm{t}\mu +
\mathrm{H}_2$ cross sections in the Ramsauer--Townsend region is
discussed elsewhere \cite{mulha99,mulha01}.

To study the sensitivity of the $\mathrm{p}\mu +\mathrm{H}_2$
scattering simulation on the cross sections we performed calculations
with the ``solid'' scattering cross sections scaled in the low energy
region (i.e., collision energy $< 0.1$~eV).
When applying a constant scaling factor, between 0.7 and 1.3, it only
changes the character of the slowing down process by a very small
amount at short times. However, it does not change the diffusion
process in any practical way.
Variations in the diffusion time $\tau_d$ and the total emission yield
do not exceed $0.5-1$\% for the mentioned scaling range.
This is not surprising since the scaling does not change the
characteristic equilibrium energy of the diffused $\mathrm{p}\mu$
which is established by the deceleration and acceleration processes
(i.e., creation and annihilations of phonons, respectively).
That energy is close to the Bragg cutoff energy $\mathrm{E}_B$,
because the coherent phonon creation for
$\mathrm{p}\mu$($\mathrm{F}=0$) vanishes below $\mathrm{E}_B$ (a weak
incoherent scattering remains) and the inelastic scattering cross
sections for the acceleration ($\sigma_{11}^{gain}$) and deceleration
($\sigma_{11}^{loss}$) cross over near $\mathrm{E}_B$
(Fig.~\ref{fig:xppp-all}).
Following these remarks, the value of $\mathrm{E}_B$ may have an
influence on the cold $\mathrm{p}\mu$ emission (but not on the
diffusion time).
However, the $\mathrm{E}_B$ value is defined by the geometrical
structure of the hydrogen crystal (lattice constant) and thus is known
relatively precisely (better than 5\%).
One result of the simulations was that such a 5\% shift of
$\mathrm{E}_B$ (0.1~meV) gives an increase (or decrease) of the
$\mathrm{p}\mu$ emission yield by 1.2\%.

Another important factor which can influence the $\mathrm{p}\mu$
diffusion parameters is the $\mathrm{pp}\mu$ molecular formation rate,
which is the most frequent $\mathrm{p}\mu$ disappearance channel
following muon decay.
Both parameters --- the diffusion time, $\tau_d$, and the emission
yield, $Y$ --- are sensitive to that process.
The value $\lambda_{pp\mu}=3.21\pm 0.18\:\mu s^{-1}$~\cite{mulha96}
used in our simulations is the more accurate value of the two existing
measurements performed in solid hydrogen to date.
Decreasing $\lambda_{pp\mu}$ by one standard deviation resulted in a
4\% increase of the calculated diffusion time and a similar increase
of the emission yield from a thick hydrogen layer.
Nevertheless, when one takes into account all above systematic errors,
the comparison presented in Tables~\ref{tab:tdif} and~\ref{tab:yields}
allows us to conclude that theory based on the ``solid'' cross section
is still consistent with the experimental results, contrary to that of
the ``gas'' approach.

In this work we tried to confront the experimental results obtained
for $\mathrm{p}\mu$ atom scattering in solid hydrogen with the theory
of low energy scattering including solid state effects.
Agreement has been obtained between experiment and that theory with
respect to $\mathrm{p}\mu$ diffusion.
In particular, the diffusion time has been found and the enhanced
yield of $\mathrm{p}\mu$ emission from the thin solid hydrogen layers
has been explained.
The results of the study proved that the observed solid effects in the
scattering at low energies (collision energy $< 0.1$~eV) are correctly
described by the calculated ``solid'' scattering cross sections.
The experimentally observed enhancement in the emission of cold
$\mathrm{p}\mu$ atoms creates the possibility of using such neutral
atomic beams as a tool in further studies of muonic processes.

\begin{acknowledgments}
The authors would like to thank V.E.~Markushin, J.L.~Douglas and
M.~Maier for fruitful discussions and for their help during the data
acquisition period.
This work is supported by the Polish Committee for Scientific
Research, the Swiss National Science Foundation, the Natural Sciences
and Engineering Research Council of Canada, TRIUMF, the Russian
Foundation for Basic Research (Grant No.~01--02--16483), and the NATO
Linkage Grant No.~LG9301162.
\end{acknowledgments}

\bibliography{mucf,ppdiff}

\begin{thebibliography}{53}
\expandafter\ifx\csname natexlab\endcsname\relax\def\natexlab#1{#1}\fi
\expandafter\ifx\csname bibnamefont\endcsname\relax
  \def\bibnamefont#1{#1}\fi
\expandafter\ifx\csname bibfnamefont\endcsname\relax
  \def\bibfnamefont#1{#1}\fi
\expandafter\ifx\csname citenamefont\endcsname\relax
  \def\citenamefont#1{#1}\fi
\expandafter\ifx\csname url\endcsname\relax
  \def\url#1{\texttt{#1}}\fi
\expandafter\ifx\csname urlprefix\endcsname\relax\def\urlprefix{URL }\fi
\providecommand{\bibinfo}[2]{#2}
\providecommand{\eprint}[2][]{\url{#2}}

\bibitem[{\citenamefont{Breunlich et~al.}(1989)\citenamefont{Breunlich, Kammel,
  Cohen, and Leon}}]{breun89}
\bibinfo{author}{\bibfnamefont{W.~H.} \bibnamefont{Breunlich}},
  \bibinfo{author}{\bibfnamefont{P.}~\bibnamefont{Kammel}},
  \bibinfo{author}{\bibfnamefont{J.~S.} \bibnamefont{Cohen}}, \bibnamefont{and}
  \bibinfo{author}{\bibfnamefont{M.}~\bibnamefont{Leon}},
  \bibinfo{journal}{Ann.~Rev.~Nucl.~Part.~Sci.} \textbf{\bibinfo{volume}{39}},
  \bibinfo{pages}{311} (\bibinfo{year}{1989}).

\bibitem[{\citenamefont{Ponomarev}(1990)}]{ponom90}
\bibinfo{author}{\bibfnamefont{L.~I.} \bibnamefont{Ponomarev}},
  \bibinfo{journal}{Contemp.~Phys.} \textbf{\bibinfo{volume}{31}},
  \bibinfo{pages}{219} (\bibinfo{year}{1990}).

\bibitem[{\citenamefont{Froelich}(1992)}]{froel92}
\bibinfo{author}{\bibfnamefont{P.}~\bibnamefont{Froelich}},
  \bibinfo{journal}{Adv.~Phys.} \textbf{\bibinfo{volume}{41}},
  \bibinfo{pages}{405} (\bibinfo{year}{1992}).

\bibitem[{\citenamefont{Measday}(2001)}]{measd01}
\bibinfo{author}{\bibfnamefont{D.}~\bibnamefont{Measday}},
  \bibinfo{journal}{Phys.~Rep.} \textbf{\bibinfo{volume}{354}},
  \bibinfo{pages}{243} (\bibinfo{year}{2001}).

\bibitem[{\citenamefont{Dzhelepov et~al.}(1965)\citenamefont{Dzhelepov,
  Ermolov, and Fil'chenkov}}]{dzhel66}
\bibinfo{author}{\bibfnamefont{V.~P.} \bibnamefont{Dzhelepov}},
  \bibinfo{author}{\bibfnamefont{P.~F.} \bibnamefont{Ermolov}},
  \bibnamefont{and} \bibinfo{author}{\bibfnamefont{V.~V.}
  \bibnamefont{Fil'chenkov}}, \bibinfo{journal}{Zh.~Eksp.~Teor.~Fiz.}
  \textbf{\bibinfo{volume}{49}}, \bibinfo{pages}{393} (\bibinfo{year}{1965}),
  \bibinfo{note}{[Sov.~Phys.~JETP {\bf 22}, 275--284 (1966)]}.

\bibitem[{\citenamefont{Bertin et~al.}(1975)\citenamefont{Bertin, Vitale, and
  Placci}}]{berti75}
\bibinfo{author}{\bibfnamefont{A.}~\bibnamefont{Bertin}},
  \bibinfo{author}{\bibfnamefont{A.}~\bibnamefont{Vitale}}, \bibnamefont{and}
  \bibinfo{author}{\bibfnamefont{A.}~\bibnamefont{Placci}},
  \bibinfo{journal}{Rivista Del Nuovo Cimento} \textbf{\bibinfo{volume}{5}},
  \bibinfo{pages}{423} (\bibinfo{year}{1975}).

\bibitem[{\citenamefont{Bertin et~al.}(1982)}]{berti82}
\bibinfo{author}{\bibfnamefont{A.}~\bibnamefont{Bertin}} \bibnamefont{et~al.},
  \bibinfo{journal}{Nuovo Cimento A} \textbf{\bibinfo{volume}{72}},
  \bibinfo{pages}{225} (\bibinfo{year}{1982}).

\bibitem[{\citenamefont{Bystritsky et~al.}(1984)}]{bystr84}
\bibinfo{author}{\bibfnamefont{V.~M.} \bibnamefont{Bystritsky}}
  \bibnamefont{et~al.}, \bibinfo{journal}{Zh.~Eksp.~Teor.~Fiz.}
  \textbf{\bibinfo{volume}{87}}, \bibinfo{pages}{384} (\bibinfo{year}{1984}),
  \bibinfo{note}{[Sov.~Phys.~JETP {\bf 60}, 219--224 (1984)]}.

\bibitem[{\citenamefont{Bracci et~al.}(1989)}]{bracc89b}
\bibinfo{author}{\bibfnamefont{L.}~\bibnamefont{Bracci}} \bibnamefont{et~al.},
  \bibinfo{journal}{Muon Catal.~Fusion} \textbf{\bibinfo{volume}{4}},
  \bibinfo{pages}{247} (\bibinfo{year}{1989}).

\bibitem[{\citenamefont{Melezhik and Wo\'{z}niak}(1992)}]{melez92}
\bibinfo{author}{\bibfnamefont{V.~S.} \bibnamefont{Melezhik}} \bibnamefont{and}
  \bibinfo{author}{\bibfnamefont{J.}~\bibnamefont{Wo\'{z}niak}},
  \bibinfo{journal}{Muon Catal.~Fusion} \textbf{\bibinfo{volume}{7}},
  \bibinfo{pages}{203} (\bibinfo{year}{1992}).

\bibitem[{\citenamefont{Adamczak et~al.}(1996)}]{adamc96}
\bibinfo{author}{\bibfnamefont{A.}~\bibnamefont{Adamczak}}
  \bibnamefont{et~al.}, \bibinfo{journal}{At.~Data and Nucl.~Data Tables}
  \textbf{\bibinfo{volume}{62}}, \bibinfo{pages}{255} (\bibinfo{year}{1996}).

\bibitem[{\citenamefont{Adamczak}(1993)}]{adamc93}
\bibinfo{author}{\bibfnamefont{A.}~\bibnamefont{Adamczak}},
  \bibinfo{journal}{Hyp.~Interact.} \textbf{\bibinfo{volume}{82}},
  \bibinfo{pages}{91} (\bibinfo{year}{1993}).

\bibitem[{\citenamefont{Cohen}(1986)}]{cohen86}
\bibinfo{author}{\bibfnamefont{J.~S.} \bibnamefont{Cohen}},
  \bibinfo{journal}{Phys.~Rev.~A} \textbf{\bibinfo{volume}{34}},
  \bibinfo{pages}{2719} (\bibinfo{year}{1986}).

\bibitem[{\citenamefont{Boukour et~al.}(1996)}]{bouko96c}
\bibinfo{author}{\bibfnamefont{A.}~\bibnamefont{Boukour}} \bibnamefont{et~al.},
  \bibinfo{journal}{Phys.~Rev.~A} \textbf{\bibinfo{volume}{53}},
  \bibinfo{pages}{3314} (\bibinfo{year}{1996}).

\bibitem[{\citenamefont{Bystritsky}(1995)}]{bystr95}
\bibinfo{author}{\bibfnamefont{V.~M.} \bibnamefont{Bystritsky}},
  \bibinfo{journal}{Nukleonika} \textbf{\bibinfo{volume}{40}},
  \bibinfo{pages}{37} (\bibinfo{year}{1995}).

\bibitem[{\citenamefont{Abbott et~al.}(1997)}]{abbot97}
\bibinfo{author}{\bibfnamefont{D.~J.} \bibnamefont{Abbott}}
  \bibnamefont{et~al.}, \bibinfo{journal}{Phys.~Rev.~A}
  \textbf{\bibinfo{volume}{55}}, \bibinfo{pages}{214} (\bibinfo{year}{1997}).

\bibitem[{\citenamefont{Marshall et~al.}(1993)}]{marsh93b}
\bibinfo{author}{\bibfnamefont{G.~M.} \bibnamefont{Marshall}}
  \bibnamefont{et~al.}, \bibinfo{journal}{Hyp.~Interact.}
  \textbf{\bibinfo{volume}{82}}, \bibinfo{pages}{529} (\bibinfo{year}{1993}).

\bibitem[{\citenamefont{Knowles et~al.}(1996)}]{knowl96}
\bibinfo{author}{\bibfnamefont{P.~E.} \bibnamefont{Knowles}}
  \bibnamefont{et~al.}, \bibinfo{journal}{Nucl.~Instrum.~Methods A}
  \textbf{\bibinfo{volume}{368}}, \bibinfo{pages}{604} (\bibinfo{year}{1996}).

\bibitem[{\citenamefont{Mulhauser et~al.}(1996)}]{mulha96}
\bibinfo{author}{\bibfnamefont{F.}~\bibnamefont{Mulhauser}}
  \bibnamefont{et~al.}, \bibinfo{journal}{Phys.~Rev.~A}
  \textbf{\bibinfo{volume}{53}}, \bibinfo{pages}{3069} (\bibinfo{year}{1996}).

\bibitem[{\citenamefont{Knowles et~al.}(1997)}]{knowl97}
\bibinfo{author}{\bibfnamefont{P.~E.} \bibnamefont{Knowles}}
  \bibnamefont{et~al.}, \bibinfo{journal}{Phys.~Rev.~A}
  \textbf{\bibinfo{volume}{56}}, \bibinfo{pages}{1970} (\bibinfo{year}{1997}),
  \bibinfo{note}{[Erratum in Phys.~Rev.~A {\bf 57}, 3136 (1998)]}.

\bibitem[{\citenamefont{Fujiwara et~al.}(2000)}]{fujiw00}
\bibinfo{author}{\bibfnamefont{M.~C.} \bibnamefont{Fujiwara}}
  \bibnamefont{et~al.}, \bibinfo{journal}{Phys.~Rev.~Lett.}
  \textbf{\bibinfo{volume}{85}}, \bibinfo{pages}{1642} (\bibinfo{year}{2000}).

\bibitem[{\citenamefont{Porcelli et~al.}(2001)}]{porce01}
\bibinfo{author}{\bibfnamefont{T.~A.} \bibnamefont{Porcelli}}
  \bibnamefont{et~al.}, \bibinfo{journal}{Phys.~Rev.~Lett.}
  \textbf{\bibinfo{volume}{86}}, \bibinfo{pages}{3763} (\bibinfo{year}{2001}).

\bibitem[{\citenamefont{Marshall et~al.}(2001)}]{marsh01}
\bibinfo{author}{\bibfnamefont{G.}~\bibnamefont{Marshall}}
  \bibnamefont{et~al.}, \bibinfo{journal}{Hyp.~Interact.}
  \textbf{\bibinfo{volume}{138}}, \bibinfo{pages}{203} (\bibinfo{year}{2001}).

\bibitem[{\citenamefont{Jacot-Guillarmod et~al.}(1996)}]{jacot96}
\bibinfo{author}{\bibfnamefont{R.}~\bibnamefont{Jacot-Guillarmod}}
  \bibnamefont{et~al.}, \bibinfo{journal}{Hyp.~Interact.}
  \textbf{\bibinfo{volume}{101/102}}, \bibinfo{pages}{239}
  (\bibinfo{year}{1996}).

\bibitem[{\citenamefont{Mulhauser et~al.}(1999)}]{mulha99}
\bibinfo{author}{\bibfnamefont{F.}~\bibnamefont{Mulhauser}}
  \bibnamefont{et~al.}, \bibinfo{journal}{Hyp.~Interact.}
  \textbf{\bibinfo{volume}{119}}, \bibinfo{pages}{35} (\bibinfo{year}{1999}).

\bibitem[{\citenamefont{Mulhauser et~al.}(2001)}]{mulha01}
\bibinfo{author}{\bibfnamefont{F.}~\bibnamefont{Mulhauser}}
  \bibnamefont{et~al.}, \bibinfo{journal}{Hyp.~Interact.}
  \textbf{\bibinfo{volume}{138}}, \bibinfo{pages}{41} (\bibinfo{year}{2001}).

\bibitem[{\citenamefont{Wo\'{z}niak et~al.}(1999)}]{wozni99}
\bibinfo{author}{\bibfnamefont{J.}~\bibnamefont{Wo\'{z}niak}}
  \bibnamefont{et~al.}, \bibinfo{journal}{Hyp.~Interact.}
  \textbf{\bibinfo{volume}{119}}, \bibinfo{pages}{63} (\bibinfo{year}{1999}).

\bibitem[{\citenamefont{Bystritsky et~al.}(2001)}]{bystr01}
\bibinfo{author}{\bibfnamefont{V.}~\bibnamefont{Bystritsky}}
  \bibnamefont{et~al.}, \bibinfo{journal}{Hyp.~Interact.}
  \textbf{\bibinfo{volume}{138}}, \bibinfo{pages}{47} (\bibinfo{year}{2001}).

\bibitem[{\citenamefont{Adamczak}(1999)}]{adamc99}
\bibinfo{author}{\bibfnamefont{A.}~\bibnamefont{Adamczak}},
  \bibinfo{journal}{Hyp.~Interact.} \textbf{\bibinfo{volume}{119}},
  \bibinfo{pages}{23} (\bibinfo{year}{1999}).

\bibitem[{\citenamefont{Gershtein}(1958)}]{gersh58}
\bibinfo{author}{\bibfnamefont{S.}~\bibnamefont{Gershtein}},
  \bibinfo{journal}{Zh.~Eksp.~Teor.~Fiz.} \textbf{\bibinfo{volume}{34}},
  \bibinfo{pages}{463} (\bibinfo{year}{1958}), \bibinfo{note}{[Sov.~Phys.~JETP
  {\bf 7}, 318 (1958)]}.

\bibitem[{\citenamefont{Vinitski{\u{\i}} and Ponomarev}(1982)}]{vinit82}
\bibinfo{author}{\bibfnamefont{S.~I.} \bibnamefont{Vinitski{\u{\i}}}}
  \bibnamefont{and} \bibinfo{author}{\bibfnamefont{L.~I.}
  \bibnamefont{Ponomarev}}, \bibinfo{journal}{Fiz.~Elem.~Chastits At.~Yadra}
  \textbf{\bibinfo{volume}{13}}, \bibinfo{pages}{1336} (\bibinfo{year}{1982}),
  \bibinfo{note}{[Sov.~J.~Part.~Nucl.~ {\bf 13}, 557--587 (1982)]}.

\bibitem[{\citenamefont{Matveenko et~al.}(1973)\citenamefont{Matveenko,
  Ponomarev, and Faifman}}]{matve75}
\bibinfo{author}{\bibfnamefont{A.~V.} \bibnamefont{Matveenko}},
  \bibinfo{author}{\bibfnamefont{L.~I.} \bibnamefont{Ponomarev}},
  \bibnamefont{and} \bibinfo{author}{\bibfnamefont{M.~P.}
  \bibnamefont{Faifman}}, \bibinfo{journal}{Zh.~Eksp.~Teor.~Fiz.}
  \textbf{\bibinfo{volume}{68}}, \bibinfo{pages}{437} (\bibinfo{year}{1973}),
  \bibinfo{note}{[Sov.~Phys.~JETP {\bf 41}, 212--216 (1973)]}.

\bibitem[{\citenamefont{Ponomarev et~al.}(1979)\citenamefont{Ponomarev, Somov,
  and Faifman}}]{ponom79}
\bibinfo{author}{\bibfnamefont{L.~I.} \bibnamefont{Ponomarev}},
  \bibinfo{author}{\bibfnamefont{L.~N.} \bibnamefont{Somov}}, \bibnamefont{and}
  \bibinfo{author}{\bibfnamefont{M.~P.} \bibnamefont{Faifman}},
  \bibinfo{journal}{Yad.~Fiz.} \textbf{\bibinfo{volume}{29}},
  \bibinfo{pages}{133} (\bibinfo{year}{1979}),
  \bibinfo{note}{[Sov.~J.~Nucl.~Phys. {\bf 29}, 67--69 (1979)]}.

\bibitem[{\citenamefont{Bubak and Faifman}(1987)}]{bubak87}
\bibinfo{author}{\bibfnamefont{M.}~\bibnamefont{Bubak}} \bibnamefont{and}
  \bibinfo{author}{\bibfnamefont{M.~P.} \bibnamefont{Faifman}},
  \bibinfo{journal}{JINR Preprint E4--87--464}  (\bibinfo{year}{1987}).

\bibitem[{\citenamefont{Struense et~al.}(1986)\citenamefont{Struense, Cohen,
  and Pack}}]{strue86}
\bibinfo{author}{\bibfnamefont{M.}~\bibnamefont{Struense}},
  \bibinfo{author}{\bibfnamefont{J.}~\bibnamefont{Cohen}}, \bibnamefont{and}
  \bibinfo{author}{\bibfnamefont{R.}~\bibnamefont{Pack}},
  \bibinfo{journal}{Phys.~Rev.~A} \textbf{\bibinfo{volume}{34}},
  \bibinfo{pages}{3605} (\bibinfo{year}{1986}).

\bibitem[{\citenamefont{Cohen and Struensee}(1991)}]{cohen91}
\bibinfo{author}{\bibfnamefont{J.~S.} \bibnamefont{Cohen}} \bibnamefont{and}
  \bibinfo{author}{\bibfnamefont{M.~C.} \bibnamefont{Struensee}},
  \bibinfo{journal}{Phys.~Rev.~A} \textbf{\bibinfo{volume}{43}},
  \bibinfo{pages}{3460} (\bibinfo{year}{1991}).

\bibitem[{\citenamefont{Melezhik et~al.}(1983)\citenamefont{Melezhik,
  Ponomarev, and Faifman}}]{melez83}
\bibinfo{author}{\bibfnamefont{V.~S.} \bibnamefont{Melezhik}},
  \bibinfo{author}{\bibfnamefont{L.~I.} \bibnamefont{Ponomarev}},
  \bibnamefont{and} \bibinfo{author}{\bibfnamefont{M.~P.}
  \bibnamefont{Faifman}}, \bibinfo{journal}{Zh.~Eksp.~Teor.~Fiz.}
  \textbf{\bibinfo{volume}{85}}, \bibinfo{pages}{434} (\bibinfo{year}{1983}),
  \bibinfo{note}{[Sov.~Phys.~JETP {\bf 58}, 254--261 (1983)]}.

\bibitem[{\citenamefont{Melezhik}(1986)}]{melez86}
\bibinfo{author}{\bibfnamefont{V.~S.} \bibnamefont{Melezhik}},
  \bibinfo{journal}{J.~Comput.~Phys.} \textbf{\bibinfo{volume}{65}},
  \bibinfo{pages}{1} (\bibinfo{year}{1986}).

\bibitem[{\citenamefont{Silvera}(1980)}]{silve80}
\bibinfo{author}{\bibfnamefont{I.~F.} \bibnamefont{Silvera}},
  \bibinfo{journal}{Rev.~Mod.~Phys.} \textbf{\bibinfo{volume}{52}},
  \bibinfo{pages}{393} (\bibinfo{year}{1980}).

\bibitem[{\citenamefont{Souers}(1986)}]{souer86}
\bibinfo{author}{\bibfnamefont{P.}~\bibnamefont{Souers}},
  \emph{\bibinfo{title}{Hydrogen Properties for Fusion Energy}}
  (\bibinfo{publisher}{University of California Press},
  \bibinfo{address}{Berkeley, California}, \bibinfo{year}{1986}).

\bibitem[{\citenamefont{Chiccoli et~al.}(1992)}]{chicc92}
\bibinfo{author}{\bibfnamefont{C.}~\bibnamefont{Chiccoli}}
  \bibnamefont{et~al.}, \bibinfo{journal}{Muon Catal.~Fusion}
  \textbf{\bibinfo{volume}{7}}, \bibinfo{pages}{87} (\bibinfo{year}{1992}).

\bibitem[{\citenamefont{Wo\'{z}niak et~al.}(1996)\citenamefont{Wo\'{z}niak,
  Bystritsky, Jacot-Guillarmod, and Mulhauser}}]{wozni96}
\bibinfo{author}{\bibfnamefont{J.}~\bibnamefont{Wo\'{z}niak}},
  \bibinfo{author}{\bibfnamefont{V.~M.} \bibnamefont{Bystritsky}},
  \bibinfo{author}{\bibfnamefont{R.}~\bibnamefont{Jacot-Guillarmod}},
  \bibnamefont{and}
  \bibinfo{author}{\bibfnamefont{F.}~\bibnamefont{Mulhauser}},
  \bibinfo{journal}{Hyp.~Interact.} \textbf{\bibinfo{volume}{101/102}},
  \bibinfo{pages}{573} (\bibinfo{year}{1996}).

\bibitem[{\citenamefont{Egelstaff}(1965)}]{egels65}
\bibinfo{author}{\bibfnamefont{P.~A.} \bibnamefont{Egelstaff}},
  \emph{\bibinfo{title}{Thermal neutron scattering}}
  (\bibinfo{publisher}{Academic Press}, \bibinfo{address}{London,New--York},
  \bibinfo{year}{1965}).

\bibitem[{\citenamefont{Fujiwara et~al.}(1997)}]{fujiw97b}
\bibinfo{author}{\bibfnamefont{M.~C.} \bibnamefont{Fujiwara}}
  \bibnamefont{et~al.}, \bibinfo{journal}{Nucl.~Instrum.~Methods A}
  \textbf{\bibinfo{volume}{395}}, \bibinfo{pages}{159} (\bibinfo{year}{1997}).

\bibitem[{\citenamefont{Marshall et~al.}(1996)}]{marsh96}
\bibinfo{author}{\bibfnamefont{G.~M.} \bibnamefont{Marshall}}
  \bibnamefont{et~al.}, \bibinfo{journal}{Hyp.~Interact.}
  \textbf{\bibinfo{volume}{101/102}}, \bibinfo{pages}{47}
  (\bibinfo{year}{1996}).

\bibitem[{\citenamefont{Olin et~al.}(1999)}]{olinx99}
\bibinfo{author}{\bibfnamefont{A.}~\bibnamefont{Olin}} \bibnamefont{et~al.},
  \bibinfo{journal}{Hyp.~Interact.} \textbf{\bibinfo{volume}{118}},
  \bibinfo{pages}{163} (\bibinfo{year}{1999}).

\bibitem[{\citenamefont{Jacot-Guillarmod}(1997)}]{jacot97b}
\bibinfo{author}{\bibfnamefont{R.}~\bibnamefont{Jacot-Guillarmod}},
  \emph{\bibinfo{title}{Stopping {C}ode}}, \bibinfo{organization}{University of
  Fribourg} (\bibinfo{year}{1997}), \bibinfo{note}{(unpublished)}.

\bibitem[{\citenamefont{Adamczak}(1997)}]{adamc97}
\bibinfo{author}{\bibfnamefont{A.}~\bibnamefont{Adamczak}},
  \emph{\bibinfo{title}{Cross Sections Data File}},
  \bibinfo{organization}{Institute of Nuclear Physics, Cracow}
  (\bibinfo{year}{1997}), \bibinfo{note}{(unpublished)}.

\bibitem[{\citenamefont{Melezhik and Wo\'{z}niak}(1998)}]{melez98}
\bibinfo{author}{\bibfnamefont{V.~S.} \bibnamefont{Melezhik}} \bibnamefont{and}
  \bibinfo{author}{\bibfnamefont{J.}~\bibnamefont{Wo\'{z}niak}},
  \bibinfo{journal}{Hyp.~Interact.} \textbf{\bibinfo{volume}{116}},
  \bibinfo{pages}{17} (\bibinfo{year}{1998}).

\bibitem[{\citenamefont{Faifman}(1989)}]{faifm89b}
\bibinfo{author}{\bibfnamefont{M.~P.} \bibnamefont{Faifman}},
  \bibinfo{journal}{Muon Catal.~Fusion} \textbf{\bibinfo{volume}{4}},
  \bibinfo{pages}{341} (\bibinfo{year}{1989}).

\bibitem[{\citenamefont{Faifman}(1988)}]{faifm88}
\bibinfo{author}{\bibfnamefont{M.~P.} \bibnamefont{Faifman}},
  \bibinfo{journal}{Muon Catal.~Fusion} \textbf{\bibinfo{volume}{2}},
  \bibinfo{pages}{247} (\bibinfo{year}{1988}).

\bibitem[{\citenamefont{Ponomarev and Faifman}(1976)}]{ponom76}
\bibinfo{author}{\bibfnamefont{L.~I.} \bibnamefont{Ponomarev}}
  \bibnamefont{and} \bibinfo{author}{\bibfnamefont{M.~P.}
  \bibnamefont{Faifman}}, \bibinfo{journal}{Zh.~Eksp.~Teor.~Fiz.}
  \textbf{\bibinfo{volume}{71}}, \bibinfo{pages}{1689} (\bibinfo{year}{1976}),
  \bibinfo{note}{[Sov.~Phys.~JETP {\bf 44}, 886--891 (1976)]}.

\bibitem[{\citenamefont{Breunlich et~al.}(1987)}]{breun87c}
\bibinfo{author}{\bibfnamefont{W.~H.} \bibnamefont{Breunlich}}
  \bibnamefont{et~al.}, \bibinfo{journal}{Muon Catal.~Fusion}
  \textbf{\bibinfo{volume}{1}}, \bibinfo{pages}{67} (\bibinfo{year}{1987}).

\end{thebibliography}

\end{document}